\documentclass[draftcls,12pt,onecolumn]{IEEEtran}
\usepackage{graphicx}
\DeclareGraphicsExtensions{.pdf,.png,.jpg}
\usepackage{amsmath}
\usepackage{latexsym}
\usepackage{graphicx}
\usepackage{bm}
\usepackage{amssymb}
\usepackage{array}
\usepackage{setspace}
\usepackage{fancyhdr}
\usepackage{citesort}
\usepackage{float}
\usepackage{algorithm}
\usepackage{algorithmicx}
\usepackage{algpseudocode} 
\newcommand{\subparagraph}{}
\usepackage{titlesec}
\usepackage[stable]{footmisc}
\usepackage{balance}
 
\usepackage{url}

\usepackage{csquotes}
\MakeOuterQuote{"}

\makeatletter
\newcommand{\vast}{\bBigg@{2.6}}
\newcommand{\Vast}{\bBigg@{4.6}}
\makeatother

\newtheorem{theorem}{Theorem}
\newtheorem{lemma}{Lemma}

\newtheorem{proposition}{Proposition}

\usepackage{parskip}

 \usepackage[
    top    = 1in,
    bottom = 1in,
    left   = 1in,
    right  = 1in]{geometry}
    
    \let\OLDthebibliography\thebibliography
\renewcommand\thebibliography[1]{
  \OLDthebibliography{#1}
  \setlength{\parskip}{0pt}
  \setlength{\itemsep}{0pt plus 0.3ex}
}

\ifCLASSINFOpdf
\else
\fi
\begin{document}
%
\title{Network-Coded Macrocell Offloading in Femtocaching-Assisted Cellular Networks}
%
%
%

\author{\IEEEauthorblockN{Yousef N. Shnaiwer${^*}$, Sameh Sorour${^*}$,  Parastoo Sadeghi$^\dagger$, Neda Aboutorab$^\dagger$, and Tareq Y. Al-Naffouri${^{*\ddagger}}$}
\IEEEauthorblockA{${^*}$Electrical Engineering Department, King Fahd University of Petroleum and Minerals (KFUPM), Dhahran, Saudi Arabia\\
$^\dagger$Research School of Engineering, The Australian National University, Canberra, Australia\\$^\ddagger$CEMSE Division, King Abdullah University of Science and Technology (KAUST), Thuwal, Saudi Arabia\\
$^*$\{g201204420,samehsorour\}@kfupm.edu.sa,\\$^\dagger$\{parastoo.sadeghi,neda.aboutorab\}@anu.edu.au,\\$^\ddagger$\{tareq.alnaffouri\}@kaust.edu.sa}}

\maketitle{}
\hfill

\begin{abstract}
The femtocaching idea was proposed as a solution to compensate for the weak backhaul capacity, by deploying coverage-limited nodes with high storage capacity called femtocaches (FCs). In this paper, the macrocell offloading  problem in femtocaching-assisted cellular networks is investigated. The objective is to minimize the number of transmissions by the macrocell base station (MBS) given that all requests should be served simultaneously to satisfy quality-of-experience (QoE) of the clients. We first formulate this MBS offloading problem as an optimization problem over a network coding graph, and show that it is NP-hard. Therefore, we propose an ONC-broadcast offloading scheme that exploits both broadcasting and opportunistic network coding (ONC) to minimize the number of required MBS transmissions. We utilize a random graph model to approximate the performance of the proposed ONC-broadcast scheme in terms of the resultant average number of transmissions by the MBS. Moreover, despite the complexity of finding the optimal solution for each and every case, we prove that this ONC-broadcast scheme is asymptotically optimal, i.e., for large number of requests,  the ONC-broadcast scheme achieves a similar  macrocell offloading performance to that of the optimal solution. To implement the ONC-broadcast scheme, we devise a heuristic that employs a dual conflict graph or broadcasting at the FCs such that the remaining requests can be served using the minimum number of transmissions at the MBS.  Simulations show that the dual graph scheme improves MBS offloading as compared to the traditional separate graph scheme. Furthermore, the simple heuristic proposed to implement the ONC-broadcast scheme achieves a very close performance to the optimal ONC-broadcast scheme.

\end{abstract}


%
\IEEEpeerreviewmaketitle

\section{Introduction}\label{IN}
Mobile data traffic is expected to increase nearly four-fold in the next three years, mainly because of video streaming \cite{CVNI}. This large demand raises the need to enhance the spectrum efficiency of next-generation cellular networks \cite{Golrezaei}. One of the recent approaches to do so in the 5G macrocell architecture is to bring the content closer to the clients, by deploying a large number of low-cost coverage-limited wireless nodes widely known as femtocells \cite{FemtoSurvey}. However, with the increase of the number of wireless nodes, the backhaul capacity becomes a bottleneck in the communication system. To alleviate this problem, a new framework has been proposed in \cite{Shanmugam}, in which these wireless nodes are equipped with large storage capacity, which is utilized to cache files according to their popularity (i.e., the files which are most likely to be requested by the clients have higher priority to be cached) \cite{Golrezaei}. These nodes are usually referred to as femtocaches (FCs) and the approach is called \emph{femtocaching} \cite{Shanmugam}. Since file popularity is slowly varying, this caching occurs through the expensive backhaul to the macrocell base station (MBS) with very low rate or during off-peak times. Clients can then download such files from the FCs at peak times. Only clients that cannot be served by the FCs at a given instant (i.e., when FCs are busy serving other clients) are allowed to download these files from the MBS to avoid undesirable video playback latency.

The success of the femtocaching framework hinges on optimizing two processes, namely, the content placement process and the content delivery process. The content placement process finds the optimum caching distribution of files or fragments of files over the FCs, so as to minimize the expected total file downloading delay \cite{Shanmugam}. A possible cost for this enhancement is having the clients download and re-assemble fragments of their requested files from multiple FCs and the MBS. On the other hand, the content delivery process  optimizes the files downloaded from each of the FCs, so as to minimize the expensive involvement (rate/bandwidth) of the MBS in this delivery process \cite{Maddah}. This paper focuses on the content delivery process assuming fixed placement of files in the caches.

Most of the solutions suggested for the content delivery problem in the literature are designed under the assumption that each client is served by only one cache \cite{Maddah,Maddah-Ali,nonuniform}. These solutions  may result in a massive number of under-utilized FCs and could be prohibitive, cost and running wise. Even when multiple clients are allowed to connect to each cache (such as in \cite{Shanmugam,Golrezaei}), it is always assumed that these clients did not previously download any of the cached files in the FCs, which contradicts the assumption about their popularity. In fact, popularity is decided based on the number of prior requests of such files from the clients. Such previously downloaded files can enhance the delivery of new files from the FCs using opportunistic network coding (ONC) \cite{Sorour}, thus further reducing the bandwidth consumed from the MBS. Indeed, ONC can exploit the diversity in prior downloaded files at the different clients to create coded combinations of currently requested files. Such combined files can be decoded at the designated clients, thus simultaneously delivering a larger number of requested files compared to simple broadcast of uncoded source files.

In this work, the content delivery problem in FC-assisted cellular networks is investigated. The diversity of the clients' file requests and prior downloads is exploited to create coding opportunities using ONC to improve various quality-of-experience (QoE) measures (e.g., delay and throughput) \cite{Markopoulou}. Our objective is to find the file coding schedule for the FCs that will minimize the bandwidth consumed by the MBS under the condition that all the requests should be served at the same time (i.e., with no scheduling latency)  to satisfy the clients' QoE. The contributions of this paper can be summarized as follows:
\begin{enumerate}
\item We formulate the macrocell offloading problem in FC-assisted cellular networks as an optimization problem over an ONC  graph, and prove its NP-hardness.
\item We propose to solve the problem in a greedy manner by utilizing either broadcasting or ONC at the FCs so as to  minimize the number of orthogonal transmissions by the MBS needed to serve the remaining requests.
\item We show that the ONC-broadcast approach is asymptotically optimal (i.e., for large number of requests, the macrocell offloading performance achieved by the ONC-broadcast scheme is similar to that of the optimal solution).
\item We analyze the performance of the aforementioned ONC-broadcast approach for a specific caching scheme using random graph theory and assuming arbitrary prior downloads and requests of files.
\item Since the ONC-broadcast scheme is also NP-hard to implement, we design a ONC-broadcast heuristic that simplifies the implementation of the ONC-broadcast scheme by utilizing well-known greedy heuristics. 
\item We elaborate on the performance of the greedy heuristic and analyze its worst-case time complexity. Furthermore, we compare the worst-case complexity of the greedy heuristic to that of the algorithm used to optimally implement the ONC-broadcast scheme (we call this algorithm the optimal ONC-broadcast scheme). Consequently, we show that the greedy heuristic reduces the worst-case complexity of implementing the ONC-broadcast scheme from an exponential to a quadratic function of the number of vertices in the MBS graph.

\end{enumerate}

The remainder of this paper is organized as follows. The related literature is summarized in Section \ref{RW}. The system model is presented in Section \ref{SM}.  The macrocell offloading problem is formulated in Section \ref{PF}. Section \ref{TGICP} introduces our proposed ONC-broadcast solution, provides an approximation of its performance, shows its asymptotic optimality, and presents the greedy heuristic. In Section \ref{SR}, simulation results are provided and discussed. Section \ref{CFW} concludes the paper.

\section{Related Work}\label{RW}

\subsection{The Caching Problem}
To the best of our knowledge, most of the literature which tackled the femtocaching problems (i.e., the content placement and the content delivery problems) have focused on the case of one cache per client and derived bounds for the throughput that can be achieved in such scenario. In \cite{Shanmugam,Dimakis}, the content placement phase is optimized under the assumptions of coded and uncoded transmissions and every client is served by one cache only. In \cite{Maddah}, the authors studied the fundamental limits of caching by introducing a scheme that maximizes a global caching gain. The same authors proposed a caching scheme for the case of isolated networks (i.e., no centrally coordinated content placement) in \cite{Maddah-Ali}, which achieves a performance close to the optimal centralized scheme. The case of non-uniform file popularity was explored in \cite{nonuniform}, and a scheme was proposed to deal with the coded caching problem by grouping files with similar popularities together. Another interesting related work is the one presented in \cite{Pedarsani}, where the online caching scenario is investigated (i.e., the file popularity is renewed at some rate according to a Markov model). The optimal online scheme was shown to approximately perform the same as the optimal offline scheme. A two-level hierarchy of caches was considered in \cite{Karamchandani}, and a scheme that provides coded multicasting opportunities within each layer and across multiple layers was devised and shown to achieve the optimal communication rates to within a constant multiplicative and additive gap. Finally, the case of multiple clients per cache was investigated in \cite{Diggavi}, and the optimal caching scheme was found for a special case (i.e., pre-determined number of caches, clients, and files). However, for the general case, each client was assumed to access only one cache during the delivery phase.

In this paper, the framework in \cite{Golrezaei}, where caching is utilized to compensate for the weak backhaul capacity, is investigated. However, our focus in this paper is on the content delivery problem, not the content placement problem as in previous works \cite{Shanmugam,TWang,Pingyod,Bahaei}. An arbitrary content placement and uniform file popularity are assumed in this paper, and the content delivery problem is solved at both the FCs and the MBS so as to offload the maximum MBS bandwidth. We call this problem the MBS offloading problem. The model assumed is general (i.e., the numbers of FCs, clients, and files are not predetermined), and a client can access any FC, and any FC can serve multiple clients (as long as their requests do not conflict with each other).

\subsection{Opportunistic Network Coding}
Network coding is a technique that allows network nodes to generate output data by encoding previously received input data \cite{Ramchandran}. Opportunistic network coding (ONC) is a network coding approach that exploits previously downloaded files at the receivers to optimize the subsequent file combinations at the transmitters \cite{DBLP:journals/corr/SorourASAA15}. The idea of ONC was first coined in \cite{COPE}, where a scheme that mixes files and broadcasts them through a wireless channel  was proposed. The scheme, named COPE, was shown to achieve significant increase in network throughput. Since then, ONC has become a popular technique for improving delay, throughput, and energy efficiency performances of wireless networks  \cite{Sorour,Seferoglu,Cao,Cui}. The most desirable property of ONC that made it so popular is the simplicity of its encoding and decoding processes which are performed using XOR operations.

The macrocell offloading problem in this paper is formulated over an ONC conflict graph, in which each vertex represents a request from a client and each edge between two vertices determines whether or not the files inducing these two vertices can be decoded  upon reception at their respective clients if XORed together. Furthermore, we utilize an ONC-based scheme, called the dual conflict graph scheme, to solve the problem at the FCs. The dual conflict graph scheme, first introduced in \cite{Habob}, uses previously downloaded files at the clients to build a dual conflict graph, which considers the transmission conflicts among the FCs (i.e., when two FCs decide to serve the same client using the same file). 

\section{System Model}\label{SM}
\subsection{Network Model}
\begin{figure}[!t]
\centering
\includegraphics[scale=.4, trim={0 2.5cm 0 2.5cm},clip]{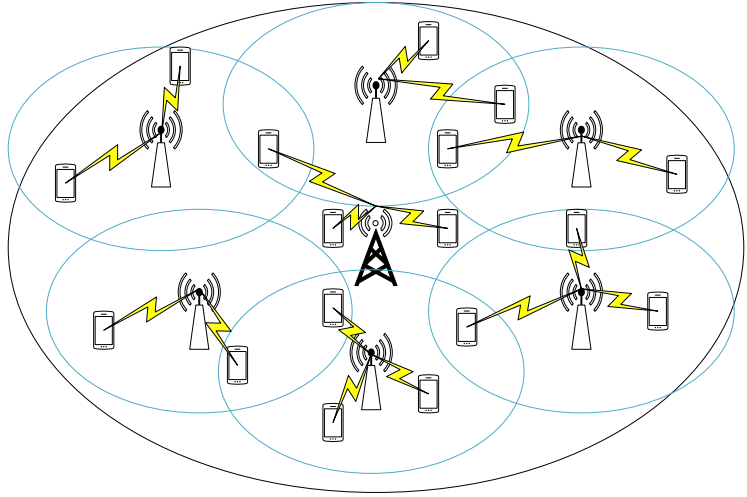}
\caption{The network model of interest where the FCs are deployed by the operator to cover the whole macrocell area. A client that cannot be served by the FCs is served by the MBS.}\label{AOD2}
\end{figure}

The network model of interest is illustrated in Fig. \ref{AOD2}. As shown in the figure, a set $\mathcal{U}=\{u_1,\dots,u_U\}$ of $U$ clients, each requesting to download/stream one file in the current time epoch  from a library $\mathcal{F}=\{f_1,\dots,f_F\}$ of $F$ popular files (i.e., the demand ratio of each client, defined as the ratio of the number of files requested by the client to the total number of files, is $\mu=\frac{1}{F}$). These files are all available at the MBS and also stored (with possible repetition) in the union of a set $\mathcal{C} = \{c_1,\dots,c_C\}$ of $C$ FCs. All the FCs are assumed to have the same coverage radius, and the coverage set of FC $c_i$ (denoted by  $\mathcal{U}(c_i)$) consists of all the clients inside the coverage radius of this FC. The set $\mathcal{H}_{c_i}$ of files stored in each FC $c_i$ is called the Has set of $c_i$, whereas the file requested by each client $u_j$ constitutes its Wants set $\mathcal{W}_{u_j}$. To guarantee clients' QoE, all requests should be served at the current time epoch, and thus all the requests that are not served by the FCs must be served by the MBS. We also assume that each client $u_j$ may have downloaded one or more files from $\mathcal{F}$ (other than the one currently in its Wants set) in previous time epochs, which constitute its Has set $\mathcal{H}_{u_j}$ at the current epoch. The FCs and the MBS can thus exploit this clients' side information to employ ONC in delivering their requests in the current epoch. It is assumed that any client has a single transceiver, and thus can only exclusively download from any one of the FCs in $\mathcal{C}$ or the MBS at a time. Moreover, it is assumed that all the files are of equal lengths.  We define the side information ratio $\sigma_{u_j}$ of client $u_j$ as the ratio of the number of files in the Has set of this client to the total number of files in the library (i.e., $H_{u_j}=|\mathcal{H}_{u_j}|=\sigma_{u_j} F,\; \forall j=1,\dots,U$). With the increase of the number of requests served by the MBS for all clients, it is assumed that the side information ratio of each client approaches the average side information ratio $\sigma_u=\frac{1}{U}\sum_{j=1}^U \sigma_{u_j}$. Additionally, we assume that the cache side information ratio $\sigma_c$, defined as the ratio of the number of files in the Has set of each FC to the total number of files in the library, is the same for all FCs (i.e., $|\mathcal{H}_{c_i}|=H_c=\sigma_c F,\; \forall i=1,\dots,C$). Finally, we assume that with proper modulation and detection methods and with channel error detection and correction techniques,  a lossless channel model can be applied in our study.

\subsection{Capabilities}
Being low-cost devices with limited wireless transmission capabilities, each FC can statically transmit over only one physical channel (e.g., a group of subcarriers in OFDMA) that is orthogonal to all the channels used by the other FCs and those of the MBS. A client scheduled to download its requested file from a specific FC must thus tune to the statically allocated channel to this FC. The MBS, being a more sophisticated wireless node, has many orthogonal channels to utilize dynamically.

Furthermore, since they are designed to be cost-efficient devices, the FCs have limited processing capabilities and thus do not participate  in solving the macrocell offloading problem (which will be formulated in Section \ref{PF}). There role is confined to sending all the relevant information to the MBS through the weak wireless backhaul, such as the requested/downloaded files by each client. All the processing is done at the MBS which has sufficient processing capabilities. In other words, the scheme proposed in this paper is fully centralized.

The MBS stores all the data received from the FCs in a log file. It then uses this data to determine which files should be combined and transmitted by the FCs and by itself (will be explained in Section \ref{TGICP}), and sends this information back to the FCs. Every FC would combine and transmit files according to the instructions received from the MBS.

\begin{figure}[!t]
\centering
\includegraphics[scale=.5]{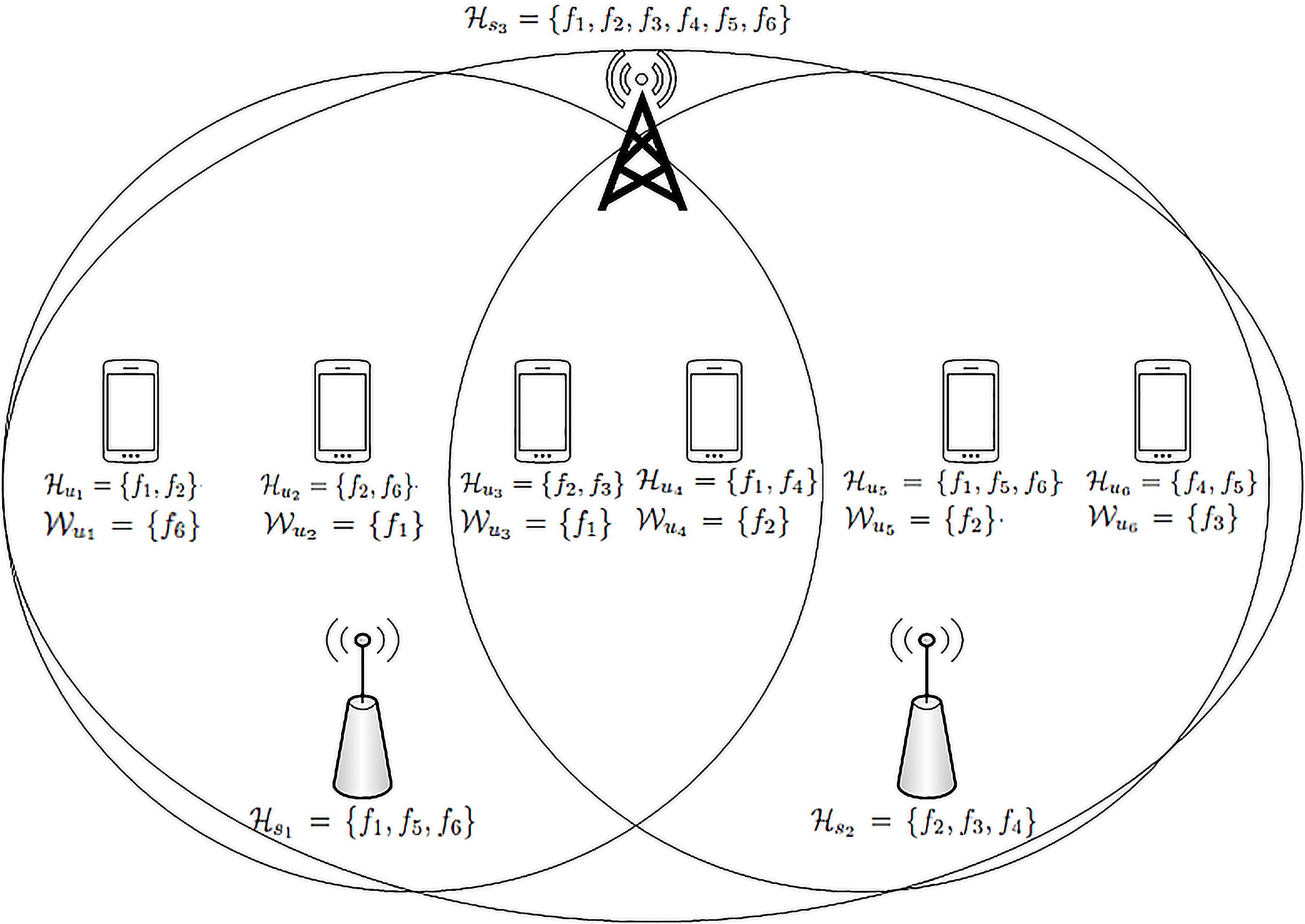}
\caption{An example of the network showing the MBS, two FCs, and six clients.}\label{Ex1}
\end{figure}

\section{Problem Formulation}\label{PF}
As mentioned in Section \ref{IN}, our ultimate target is to minimize the number of orthogonal channels needed from the MBS to serve the requests that are not served by the FCs. In general, when the files are requested by the clients, each of the FCs can  transmit a combination of a subset of its cached  files that could be decoded by their requesting clients. The question now is: \emph{Which coded/uncoded files should be transmitted by each of the FCs, such that the remaining requested files (if any) can be delivered  to the remaining clients using the minimum number of orthogonal MBS channels? }

\subsection{Motivating Example} \label{sec:me}
To illustrate the problem of interest in this paper, let us consider the network scenario shown in Fig. \ref{Ex1}. The example encompasses the MBS with its Has set, which constitutes the whole library, two FCs with their Has sets collectively storing the whole library, and six clients with their Has and Wants sets. The MBS is denoted as if it was a third FC to simplify the notation.

In this example, several schedules of coded/uncoded file downloads can be envisaged. We will focus on two solutions. In what follows, $f_{k_1}\oplus f_{k_2}$ is an XOR combination of the bits of files $f_{k_1}$ and $f_{k_2}$.\\\\
\textbf{Solution 1}:
\begin{itemize}
\item FC $c_1$ transmits $f_1\oplus f_6$: This allows $u_1$ and $u_2$ to decode their requested files as they already have $f_1$ and $f_6$, respectively.
\item FC $c_2$ transmits $f_2$, which satisfies the requests of $u_4$ and $u_5$.
\end{itemize}
In this case, the MBS must simultaneously serve $u_3$ and $u_6$ requesting files $f_1$ and $f_3$, respectively. Since $u_6$ does not have $f_1$, an XOR of $f_1 \oplus f_3$ cannot be decoded at $u_6$. To satisfy the no-latency constraint, the MBS must thus transmit $f_1$ and $f_3$ uncoded on separate orthogonal channels to $u_3$ and $u_6$, respectively. Consequently, this solution will end up consuming two channels from the MBS.  \\\\
\textbf{Solution 2}:
\begin{itemize}
\item FC $c_1$ transmits $f_1\oplus f_6$, which addresses the requests of $u_1$ and $u_2$ as in the previous solution.
\item FC $c_2$ transmits $f_3$, which satisfies the requests of $u_6$.
\end{itemize}
In this case, the MBS must simultaneously serve the requests of $u_3$, $u_4$ and $u_5$ requesting files $f_1$, $f_2$ and $f_2$, respectively. By looking at the Has sets of these clients, we can clearly see that they can all decode their requested files from a single coded transmission of $f_1\oplus f_2$. Thus, this solution will end up consuming only one channel from the MBS, which is half the bandwidth required by Solution 1.\\

By searching over all the possible options, we can see that Solution 2 (though not being the unique optimal solution as will be discussed later) results in the minimum number of consumed channels at the MBS. Now the interesting question is how we can systematically find the optimal solution(s) for any scenario with any size of network. In the next section, we formulate this problem as an optimization problem over an ONC graph.

\subsection{Graph-Based Formulation}\label{sec:FOR}
To formulate the above problem, we first need to define the ONC conflict graph $\mathcal{G}$ that represents all the possible coding conflicts (i.e., files that when XORed together cannot be immediately decoded at their requesting clients). This ONC conflict graph is constructed as follows. Every client $u_j$, requesting file $f_k$, has only one vertex $v_{j,k}$ in the graph. Two vertices $v_{j_1,k_1}$ and $v_{j_2,k_2}$ will be set adjacent by an edge in this graph if they represent the request of two different files while at  least one of the clients inducing the two vertices does not possess the file requested by the other client  (i.e., $f_{k_1}\neq f_{k_2}$ AND either $f_{k_1} \notin \mathcal{H}_{c_{j_2}}$ OR  $f_{k_2} \notin \mathcal{H}_{c_{j_1}}$). Therefore, at least one of the two clients $u_{j_1}$ or $u_{j_2}$ will not be able to extract its own requested file from an XOR of $f_{k_1} \oplus f_{k_2}$.

\begin{figure}[!t]
\centering
\includegraphics[scale=.5]{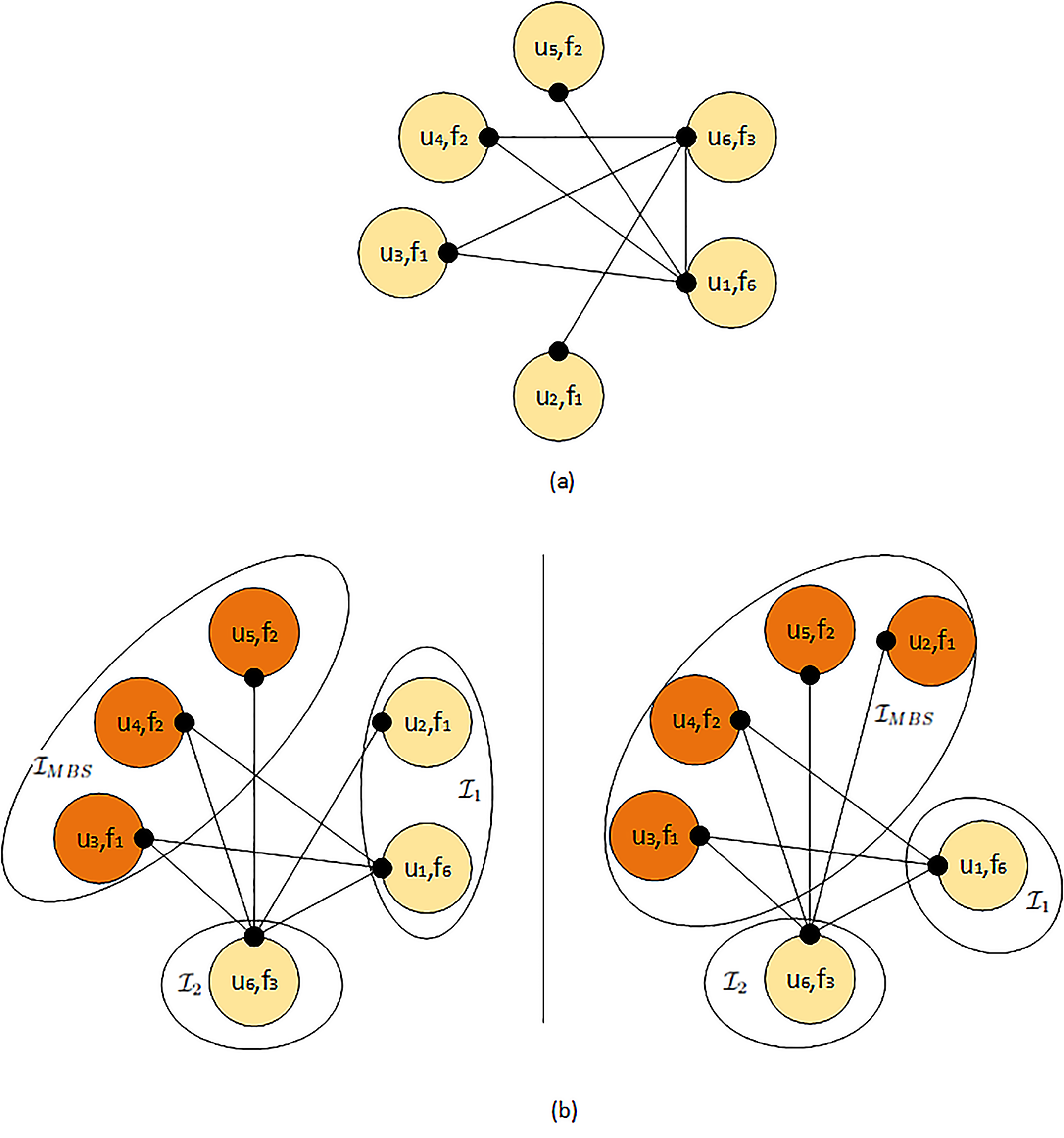}
\caption{(a) An illustration of the ONC conflict graph for the scenario in Fig. \ref{Ex1}. (b) Two possible optimal solutions for the given scenario.}\label{FOR}
\end{figure}

Fig. \ref{FOR}.a depicts the ONC conflict graph of the example in Fig. \ref{Ex1}.  Only adjacent vertices based on the above conditions cannot be encoded with each other. Thus, defining any independent set\footnote{An independent set in a graph is a set of pairwise non-adjacent vertices.} $\mathcal{I}$ in this graph, the XOR of all files represented in the vertices of $\mathcal{I}$ (which will be denoted by $\mathcal{F}(\mathcal{I})$) can be decoded immediately by all clients represented in the vertices of $\mathcal{I}$ (which will be denoted by $\mathcal{U}(\mathcal{I})$). Thus, every independent set in graph $\mathcal{G}$ can consume the channel of an FC or one of the MBS channels.

Now, let $\mathcal{I}_i,\; i=1,\dots,C$, be the independent set, whose files $\mathcal{F}(\mathcal{I}_i)$ will be combined and transmitted by FC $c_i$ to satisfy the requests of the clients in $\mathcal{U}(\mathcal{I}_i)$. Our problem is then reduced to finding the set of independent sets $\mathcal{I}_1, \dots, \mathcal{I}_C$, so as to minimize the remaining number of independent sets in $\mathcal{G}'=\mathcal{G}\setminus \bigcup_{i=1}^C \mathcal{I}_i$. Indeed, these remaining independent sets must be served by orthogonal channels at the MBS, and thus the target is to minimize their number. Since the minimum number of independent sets in graph $\mathcal{G}'$ is equal to its chromatic number\footnote{The chromatic number of a graph is the minimum number of colors, with which the vertices of the graph could be colored, such that no two adjacent vertices carry the same color.} $\chi(\mathcal{G}')$ \cite{opac-b1130915}, we can thus formulate our problem as

\small
\begin{equation}\label{Eq.17}
\begin{aligned}
& \min_{\mathcal{I}_1,\dots, \mathcal{I}_C} \;\;\; \chi\left(\mathcal{G}\backslash \bigsqcup_{i=1}^C \mathcal{I}_i\right)\\
& \mathrm{subject \; to} \;\;  \mathcal{F}(\mathcal{I}_i)\subseteq \mathcal{H}_{c_i},\; \forall c_i\in\mathcal{C},\\
& \;\;\;\;\;\;\;\;\;\;\;\;\;\;\;\;\; \mathcal{U}(\mathcal{I}_i)\subseteq \mathcal{U}(c_i), \;\; \forall c_i\in\mathcal{C},
\end{aligned}
\end{equation}
\normalsize

where $\bigsqcup$ is the disjoint union operator and indicates that the independent set of each FC should be disjoint from all the other independent sets. The first constraint in (\ref{Eq.17}) ensures that all the files to be served by the $i^{th}$ FC exist in its \textbf{Has} set. The second constraint indicates that every client whose request is included in the independent set $\mathcal{I}_i$ should be in the coverage set of the $i^{th}$ FC. For instance, if the first FC in the example shown in Fig. \ref{Ex1} served clients 2, 3, 4 and 5 by XORing and transmitting the files $f_1$ and $f_2$ and the second FC served the last client by  transmitting the file $f_3$, then the MBS needs only one transmission to serve the first client. However, the first FC does not have file $f_2$ in its \textbf{Has} set (i.e., $f_2\notin \mathcal{H}_{c_1}$), and client 5 is not in the coverage radius of the first FC (i.e., $u_5\not\subseteq \mathcal{U}(c_1)$). Consequently, this is not a valid solution to the problem in (\ref{Eq.17}) because the constraints are not satisfied.

The left subfigure of Fig. \ref{FOR}.b depicts one optimal selection of $\mathcal{I}_1$ and $\mathcal{I}_2$ for the example in Fig. \ref{Ex1}, resulting in Solution 2 (explained in Section \ref{sec:me}) and a chromatic number of 1 for the remaining vertices. The right subfigure shows another selection of $\mathcal{I}_1$ and $\mathcal{I}_2$ achieving the same chromatic number, which means that more than one optimal solution may exit for the same scenario. In the following theorem, we show that the problem in (\ref{Eq.17}) is NP-hard.

\begin{theorem}\label{Thm5}
The macrocell offloading problem in (\ref{Eq.17}) is NP-hard.
\end{theorem}

\begin{IEEEproof}
We prove the theorem using polynomial time Turing machine reduction \cite{Ladner}. First, we recall that our problem is to find the maximal independent set, among all maximal independent sets in the conflict graph, that minimizes the number of MBS transmissions when optimal coloring is utilized at the MBS. Turing machine reduction works as follows. The part of the problem that is known to be NP-hard is put in a subroutine. If solving the problem requires calling this subroutine polynomial number of times, then the problem is polynomial-time reduced. Since the problem solved by the subroutine is NP-hard, then the whole problem is NP-hard.

Therefore, the first question to be asked is: which part of the problem in (\ref{Eq.17}) is known to be NP-hard to solve? Clearly, the answer is the optimal graph coloring at the MBS. We assume a small number of vertices ($\nu<5$) and we show that the problem is NP-hard for this number. Then the problem would be NP-hard also for $\nu>=5$, since the complexity grows with the number of vertices.

For $\nu<5$, the maximum number of maximal independent sets in a graph is equal to $\nu$ \cite{Jerrold}. Thus, the optimal graph coloring subroutine is called $\nu$ times at most, which is polynomial in $\nu$. Hence, the problem in (\ref{Eq.17}) is NP-hard.
\end{IEEEproof}

\begin{figure}[!t]
\centering
\includegraphics[scale=.6, trim={3cm 9cm 3cm 9cm},clip]{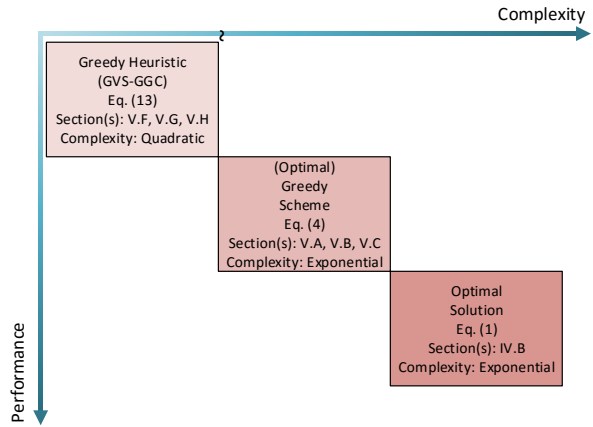}
\caption{Performance-complexity comparison of the different solutions of the MBS offloading problem.}\label{PCD}
\end{figure}

Different methods of solving the MBS offloading problem in (\ref{Eq.17}) are discussed in this paper, and the performance and complexity of these methods are compared in Fig. \ref{PCD}. The optimal solution is the one obtained by solving (\ref{Eq.17}) exactly. The ONC-broadcast scheme (to be discussed in detail in Section \ref{TGICP}) limits the number of possible schemes to choose from at the MBS to four schemes based on whether the FCs and the MBS use ONC or uncoded broadcasting to transmit the files. The greedy vertex search-greedy graph coloring (GVS-GGC), or the greedy heuristic, is used to simplify the implementation of the ONC-broadcast scheme, and will be explained in Section \ref{GS}. 

\section{The Proposed ONC-Broadcast Scheme}\label{TGICP}
In this Section, a ONC-broadcast scheme is proposed to solve the problem in (\ref{Eq.17}). The rationale behind this scheme is  described and explained first. Next, the asymptotic optimality of this ONC-broadcast scheme is shown. After that, the performance of the ONC-broadcast scheme is analyzed in terms of the number of required MBS transmissions.  Finally, a simple and efficient heuristic is proposed to implement the ONC-broadcast scheme.

\subsection{The ONC-Broadcast Scheme Philosophy}\label{TGSP}
The philosophy of our proposed ONC-broadcast scheme is to utilize either uncoded broadcasting or ONC at the FCs, such that the required number of transmissions at the MBS can be minimized. In other words, the ONC-broadcast scheme simplifies the process of solving (\ref{Eq.17}) at the MBS to choosing, among the four following schemes, the one that achieves the minimum number of MBS transmissions:
\begin{enumerate}
\item The FCs broadcast $C$ uncoded files (out of $F$ files in the library), and the MBS broadcasts the remaining $F-C$  files in an uncoded manner.
\item The FCs broadcast $C$ uncoded files, and the MBS uses an ONC conflict graph to serve the remaining requests.
\item The FCs use an ONC conflict graph\footnote{The conflict graph formed by the FCs is a bit different than the one described in Section \ref{sec:FOR}, and would be explained in Section \ref{HBSACG}.} to transmit $C$ coded/uncoded files, and the MBS broadcasts the files which are not completely served by the FCs in an uncoded manner.
\item The FCs exploit an ONC conflict graph to transmit $C$ coded/uncoded files, and the MBS utilizes an ONC conflict graph to serve the remaining requests.
\end{enumerate} 
Using ONC at the FCs aims at maximizing the number of requests served by them, which is equivalent to minimizing the number of requests that should be served by the MBS, with the expectation that this minimum number of remaining requests could be served in the minimum number of orthogonal channels. One may think that this is intuitively true in general. However, we can easily show that utilizing ONC alone at the FCs does not always result in the minimum number of orthogonal MBS channels as compared to uncoded broadcasting. 

\begin{figure}[!t]
\centering
\includegraphics[scale=.5]{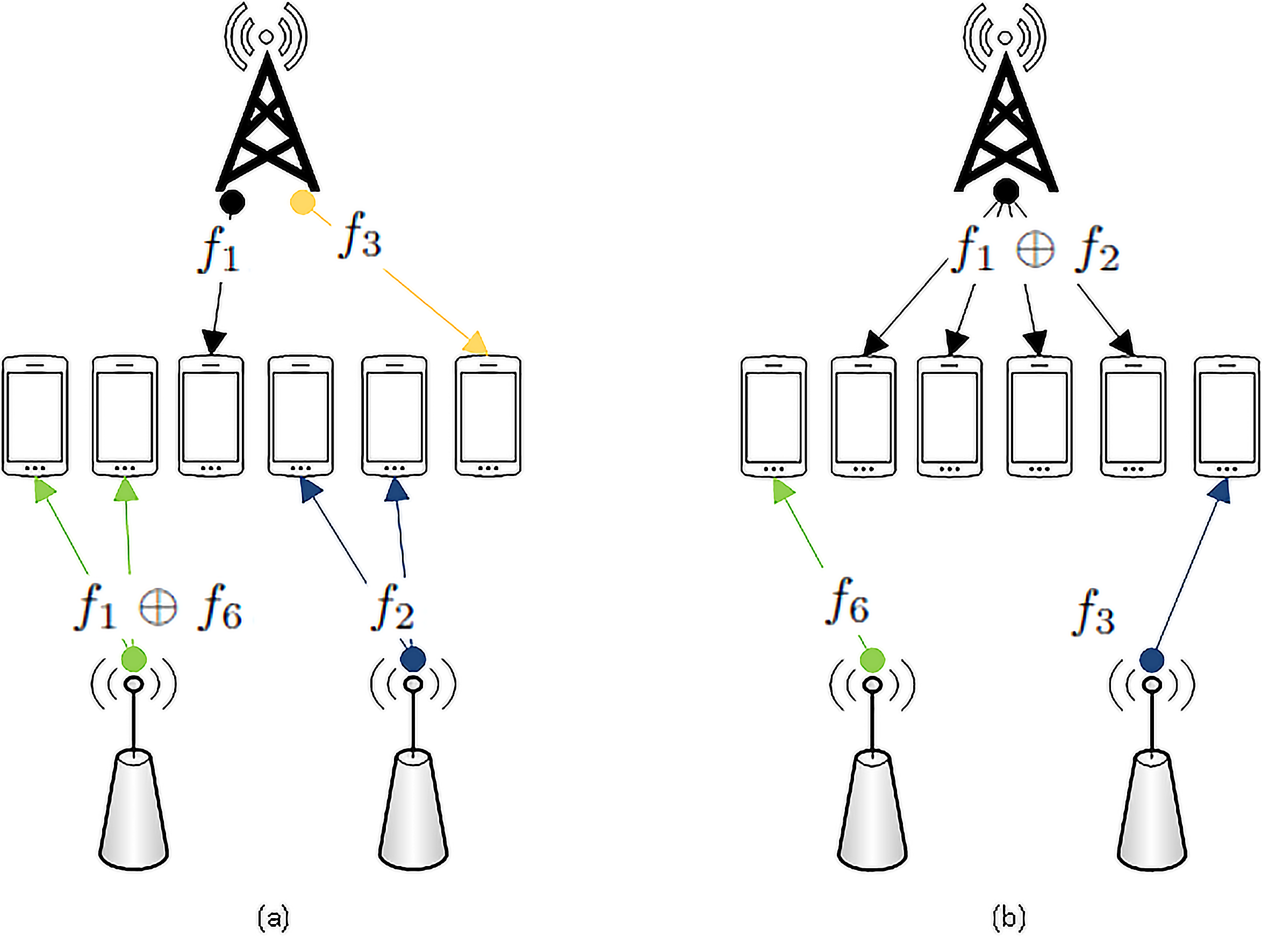}
\caption{Two examples of the FC solution. In (a), the ONC solution results in two transmissions by the MBS for two requests, but in (b), the broadcast solution needs only one MBS transmission to serve the four remaining requests.}\label{Ex}
\end{figure}

To get better understanding of this fact, take the example illustrated in Fig. \ref{Ex}, which shows two different possible solutions of the problem in (\ref{Eq.17}) for the same model of Fig. \ref{Ex1}. The solution shown in Fig. \ref{Ex}.a is the one that maximizes the number of requests to be served by the FCs using ONC. However, this solution is not optimal, because it results in two orthogonal transmissions by the MBS. One optimal solution to the same scenario is shown in Fig. \ref{Ex}.b, which requires only one transmission by the MBS although it serves more requests (i.e., less requests are served by the FCs). This optimal solution  employs uncoded broadcasting  at the FCs.

\subsection{Mathematical Model}

In this section, we provide a mathematical description of the proposed ONC-broadcast scheme. Based on the demonstration of the ONC-broadcast scheme in Section \ref{TGSP}, the number of MBS transmissions can be written as 

\small
\begin{equation}\label{Eq.200}
\begin{aligned}
N_{MBS} &=min(F-C,\chi(U-N_{FC}^{(B)}),F-F_B,\chi(U-N_{FC}))=min(F-C,\chi(U-N_{FC}^{(B)}),\chi(U-N_{FC})),
\end{aligned}
\end{equation}
\normalsize

where $N_{FC}^{(B)}$ is the number of served clients when uncoded broadcasting is utilized at the FCs, $F_B$ is the number of broadcast uncoded files in the FCs' graph solution, $N_{FC}$ is the number of served clients when the FCs use ONC to transmit. The four terms inside the $min()$ function in the first inequality of (\ref{Eq.200}) respectively correspond to the four possible schemes described in Section \ref{TGSP}.  The second equality in (\ref{Eq.200}) follows from the fact that $C$ is the maximum number of uncoded files that can be broadcast by the FCs, thus $F-C\leq F-F_B$. Next, we simplify (\ref{Eq.200}) further by using the following theorem.

\begin{theorem}\label{Thm6}
The number of clients served when the FCs use ONC is lower bounded by the number of clients served by the FCs when uncoded broadcasting is utilized, i.e.,
\begin{equation}
N_{FC}^{(B)}\leq N_{FC}.
\end{equation}
\end{theorem}

\begin{IEEEproof}
See Appendix A for the proof of this Theorem.
\end{IEEEproof}

From Theorem \ref{Thm6}, it can be inferred that $\chi(U-N_{FC}^{(B)})\geq \chi(U-N_{FC})$ \footnote{This conclusion is based on the fact that the average chromatic number of a graph is monotonically increasing with the number of vertices, which is formally proven in Theorem \ref{Thm2} (Section \ref{OGS}) using random graph theory. However, this fact can be intuitively shown by iteratively adding vertices to a random graph and observing that the average number of colors used to cover all the vertices is increasing with the addition of every new vertex.}. Moreover, since the chromatic number of a graph is always less than or equal to the number of vertices in the graph ($\chi(U-N_{FC})\leq U-N_{FC}$), then we can break $N_{MBS}$ in (\ref{Eq.200}) into two cases as follows:

\small
\begin{equation}\label{Eq.201}
N_{MBS}=
    \begin{cases}
      \chi(U-N_{FC}), &  U-N_{FC}\leq F-C \\
      \min(\chi(U-N_{FC}),F-C), & U-N_{FC}> F-C
    \end{cases}.
\end{equation}
\normalsize

The interpretation of (\ref{Eq.201}) is as follows. The minimum number of transmissions from the MBS depends on the relation between the number of remaining files to be served by the MBS if the FCs used uncoded broadcast ($F-C$) and the number of remaining clients to be served by the MBS if the FCs used ONC ($U-N_{FC}$). If $U-N_{FC} \leq F-C$, then the minimum number of MBS transmissions is achieved if the FCs use ONC to serve $N_{FC}$ clients and the MBS also uses ONC to serve the remaining clients using $\chi(U-N_{FC})$ channels. However, if $U-N_{FC} > F-C$, it is not immediately clear which scheme is better and the MBS has to compare the chromatic number of its resulting graph $\chi(U-N_{FC})$ with its uncoded transmissions $F-C$ to decide whether both FCs and MBS should uses ONC or both should utilize uncoded broadcasting.

\subsection{Asymptotic Optimality of the ONC-Broadcast Solution}\label{OGS}
In this part, we show that the proposed ONC-broadcast scheme is asymptotically optimal, i.e., with the increase of the number of  the clients' requests served by the MBS, the performance of the ONC-broadcast scheme approaches that of the optimal solution.  To do so, we need to show that $N_{MBS}$ in (\ref{Eq.201}) is a non-decreasing function on the number of clients $U$, which implies that minimizing the remaining number of clients in the MBS graph results in the minimum number of orthogonal transmissions needed by the MBS. 

We start by modeling the MBS  conflict graph $\mathcal{G}^c$ using a random graph model. The new random MBS conflict graph is defined as  $\mathcal{G}^c_{\nu,\pi}$, where $\nu=U-N_{FC}$ denotes the number of vertices, and $\pi$ is the probability that two vertices are connected in the MBS conflict graph. 

Finding the average chromatic number of a random graph depends on determining the connectivity probability $\pi$, and the number of vertices $\nu$. One lemma proved in \cite{Bollobas} can be used to approximate the chromatic number of $\mathcal{G}^c_{\nu,\pi}$ as

\small
\begin{equation}\label{Eq.3}
\chi(\mathcal{G}^c_{\nu,\pi})=\left(\frac{1}{2}+o(1)\right)\log\left(\frac{1}{1-\pi}\right)\frac{\nu}{\log(\nu)}.
\end{equation}
\normalsize

In the following lemma, the connectivity probability $\pi$ is derived as a function of $\nu$.
\begin{lemma}\label{Lemma1}
Given $\nu$, the probability of having two vertices connected in $\mathcal{G}^c$ is expressed as

\small
\begin{equation}\label{Eq.10}
\pi=(1-\sigma_u^2)\frac{\nu(F-1)}{\nu F-1}.
\end{equation}
\end{lemma}
\normalsize

\begin{IEEEproof}
See Appendix B for the proof of this lemma.
\end{IEEEproof}

\begin{theorem}\label{Thm2}
The ONC-broadcast scheme is asymptotically optimal.
\end{theorem}

\begin{IEEEproof}
By substituting (\ref{Eq.10}) in (\ref{Eq.3}), the chromatic number of $\mathcal{G}$ as a function of $\nu=U-N_{FC}$ can be expressed as

\small
\begin{equation}\label{Eq.16}
\begin{aligned}
\chi(\mathcal{G}_{\nu,\pi})&=\left(\frac{1}{2}+o(1)\right)\log\left(\frac{1}{1-(1-\sigma_u^2)\frac{\nu  (F-1)}{\nu F-1}}\right)\frac{\nu}{\log(\nu)}&\\
& =\left(\frac{1}{2}+o(1)\right)\log\left(\frac{\nu F-1}{(\nu F-1)-((1-\sigma_u^2)\nu (F-1))}\right)  \frac{\nu}{\log(\nu)}.
\end{aligned}
\end{equation}
\normalsize

Clearly the term $((1-\sigma_u^2)\nu (F-1))$ increases with $\nu$, so the ratio of the numerator to the  denominator in the term inside the logarithm increases with $\nu$. Thus, the log term is a monotonically increasing function of $\nu$. Since $\frac{\nu}{\log(\nu)}$ is increasing with $\nu$ ($\nu>\log(\nu)$), we conclude that the chromatic number of $\mathcal{G}$ is increasing with $\nu$. This implies that $N_{MBS}$ in (\ref{Eq.201}) is non-decreasing with $\nu$. Hence, the ONC-broadcast scheme is asymptotically optimal.
\end{IEEEproof}

\subsection{Maximizing File Downloads from FCs}\label{HBSACG}
The intuitive way of implementing the ONC-broadcast scheme at the FCs is to allow each FC to serve the maximum number of clients by building and solving its own conflict graph independently from other FCs. However, this separate graph method does not consider service conflicts (i.e., when two FCs decide to serve the same client), and this weakens the MBS offloading capabilities of the FCs \cite{Shnaiwer}.  To avoid such service conflicts, we will use a dual conflict ONC graph, which was initially introduced in \cite{Habob} and can be described as follows. For every client $u_j$ requesting file $f_k$, we will generate a set of vertices $v_{i,j,k}$ $\forall~c_i$ having $f_k \in \mathcal{H}_{c_i}$ and $u_j\in\mathcal{U}(c_i) $. In other words, the requested file $f_k$ by client $u_j$ is represented by several vertices in the graph, each representing an FC having $f_k$ in its cache and $u_j$ in its coverage set, which means it can address the request of $u_j$. Now any two vertices $v_{i_1,j_1,k_1}$ and $v_{i_2,j_2,k_2}$ will be set adjacent by an edge if one of the following conditions occur:
\begin{enumerate}\label{cond_2}
\item  $f_{k_1}\neq f_{k_2}$ AND either $f_{k_1} \notin \mathcal{H}_{c_{j_2}}$ OR  $f_{k_2} \notin \mathcal{H}_{c_{j_1}}$.
\item  $j_1 = j_2$ AND $i_1\neq i_2$.
\end{enumerate}

The first condition is clearly the same as the coding conflict condition of the original ONC graph described in Section \ref{sec:FOR}. The second condition adds  service conflict edge between any two vertices representing the service of the same client by two different FCs. With this structure, it can be easily inferred that any independent set in this dual conflict graph will represent a full schedule of coded file downloads to the FCs without any two FCs trying to download simultaneously to the same client.

\subsection{Performance Analysis}\label{PA}
In this subsection, we derive an approximation of the average number of MBS transmissions for the proposed ONC-broadcast scheme. To simplify the analysis in this part, we assume systematic fixed placement process is used to distribute the files among the FCs, which can be described as follows. Assuming that each FC can store $F_c$ files, the first $F_c$ files in the library are stored in the first FC, then the second $F_c$ files are stored in the second FC, and so on until all the FCs' caches are filled. If the last file in the library is reached before all the caches are filled, the first file is filled next and so on.

For large FC cache size (which is usually the case), each file is guaranteed to be repeatedly stored among the FCs. We define $R=\frac{H_c C}{F}$ as the repetition index which represents the number of copies of each file in the caches of all FCs. All the derivations in this section are based on the assumption that $R$ takes integer values only (this would limit the use of the derived expressions to integer values of $R$, but it does not affect their accuracy). We also define $B=C/R$ as the number of FCs having all the files.

In the following theorem, we derive approximate expressions for both $N_{FC}$ and $\chi(U-N_{FC})$ in (\ref{Eq.201})  by employing random graph theory to model both the FC dual conflict graph and the MBS conflict graph. 

\begin{theorem}\label{Thm3}
When systematic fixed placement is utilized to distribute the files among the FCs, the chromatic number of the MBS graph $\chi(U-N_{FC})$ is given by

\footnotesize
\begin{equation}\label{Eq.40}
\begin{aligned}
 \chi(U-N_{FC})=& \sum_{\tilde{\nu}=2}^{RU} \left(\frac{1}{2}+o_2(1)\right)\log\left(\frac{(U-N_{FC|\tilde{\nu}})F-1}{((U-N_{FC|\tilde{\nu}})F-1)-(1-\sigma_u^2)(U-N_{FC|\tilde{\nu}})(F-1)}\right)\times &\\
&\frac{}{} \frac{(U-N_{FC|\tilde{\nu}})\binom{RU}{\tilde{\nu}}(\sigma_c P_{\mathcal{C}})^{\tilde{\nu}}(1-\sigma_c P_{\mathcal{C}})^{RU-\tilde{\nu}}}{\log(U-N_{FC|\tilde{\nu}})(1-(1-\sigma_c P_{\mathcal{C}})^{RU-1}(\sigma_c P_{\mathcal{C}}(RU-1)+1))},
\end{aligned}
\end{equation}
\normalsize

where $N_{FC|\tilde{\nu}}=2\log_{\frac{1}{1-\tilde{\pi}}}\left(\frac{e\tilde{\nu}}{2\log_{\frac{1}{1-\tilde{\pi}}}(\tilde{\nu})}\right)+1+o_1(1)$, $P_{\mathcal{C}}$ is the probability that a client is in the coverage set of an FC, and $\tilde{\pi}$ is expressed as
\end{theorem}

\small
\begin{equation}\label{Eq.130}
\begin{aligned}
& \tilde{\pi}= \sum_{\overset{y_v=0}{\forall v\neq C}}^U \sum_{y_C=\max(0,2-\sum_{q=1}^{C-1}y_q)}^U \sum_{m=1}^C  \frac{y_m^2(y_m-1)(H_c-1)\prod_{v=1}^{C}\binom{U}{y_v} (P_{\mathcal{C}})^{y_v} (1-P_{\mathcal{C}})^{U-y_v}}{\sum_{p=1}^C y_p(\sum_{p=1}^C y_p-1)(y_m H_c-1)\left(1-(1-P_{\mathcal{C}})^{UC-1}(P_{\mathcal{C}}(UC-1)+1)\right) }  \times  \\
& \vast(1-\sum_{e'_1=0}^{H_u}\frac{\binom{H_u}{e'_1} \binom{F-H_u}{H_u-e'_1}}{\binom{F}{H_u}} \sum_{e'_2=\max(0,H_u-(F-H_c))}^{\min(H_u-e'_1,H_c)}\frac{\binom{H_c}{e'_2} \binom{F-H_c}{H_u-e'_1-e'_2}}{\binom{F}{H_u-e'_1}} \sum_{e'_3=\max(0,H_u-(F-(H_c-e'_2)))}^{min(H_u-e'_1,H_c-e'_2)} \frac{\binom{H_c-e'_2}{e'_3} \binom{F-(H_c-e'_2)}{H_u-e'_1-e'_3}}{\binom{F}{H_u-e'_1}} \frac{e'_2}{H_c}\frac{e'_3}{H_c-1}\vast)\\
& +\frac{R-1}{UR-1}.
\end{aligned}
\end{equation}
\normalsize

\begin{IEEEproof}
See Appendix C for the proof of this Theorem.
\end{IEEEproof}

Before we end this section, we derive $N_{FC}$ and $\chi(U-N_{FC})$ for an interesting special case, namely, when the coverage set of each FC contains all clients. The closest practical example to this scenario is the model investigated in \cite{Radaydeh1,Radaydeh}, where a group of femtocell access points are deployed in a public area (e.g., metro stations) to improve data rate and coverage. In our case, the femtocell access points are replaced with femtocaches to enhance macrocell offloading. There are three main differences between this scenario and the general case. First, the coverage set of each FC contains all the clients. Second, the number of vertices induced by each client is fixed and is equal to $R$, since a client requesting a file induces a number of vertices equal to the number of copies of this file in all FCs. The last difference is that the number of vertices in the FCs' dual conflict graph $\tilde{\nu}$ is not random any more but a fixed value $\tilde{\nu}=\mu UCF\sigma_c$.

By applying the aforementioned changes to the proof in Appendix C, it can be shown that the number of transmissions by the MBS for the full FC coverage is expressed as

\small
\begin{equation}\label{Eq.500}
\begin{aligned}
& \chi(U-N_{FC})= \left(\frac{1}{2}+o_2(1)\right)\log\left(\frac{(U-N_{FC})F-1}{((U-N_{FC})F-1)-(1-\sigma_u^2)(U-N_{FC})(F-1)}\right) \frac{U-N_{FC}}{\log(U-N_{FC})},
\end{aligned}
\end{equation}
\normalsize
where $N_{FC}=2\log_{\frac{1}{1-\tilde{\pi}}}\left(\frac{e\tilde{\nu}}{2\log_{\frac{1}{1-\tilde{\pi}}}(\tilde{\nu})}\right)+1+o_1(1)$, and $\tilde{\pi}$ is written as

\small
\begin{equation}\label{Eq.501}
\begin{aligned}
& \tilde{\pi}=\frac{R(R-1)}{\tilde{\nu}(\tilde{\nu}-1)}+ \frac{U(U-1)(H_c-1)}{(UC-1)(U H_c-1)}  \times  \\
& \vast(1-\sum_{e'_1=0}^{H_u}\frac{\binom{H_u}{e'_1} \binom{F-H_u}{H_u-e'_1}}{\binom{F}{H_u}} \sum_{e'_2=\max(0,H_u-(F-H_c))}^{\min(H_u-e'_1,H_c)}\frac{\binom{H_c}{e'_2} \binom{F-H_c}{H_u-e'_1-e'_2}}{\binom{F}{H_u-e'_1}} \sum_{e'_3=\max(0,H_u-(F-(H_c-e'_2)))}^{min(H_u-e'_1,H_c-e'_2)} \frac{\binom{H_c-e'_2}{e'_3} \binom{F-(H_c-e'_2)}{H_u-e'_1-e'_3}}{\binom{F}{H_u-e'_1}} \frac{e'_2}{H_c}\frac{e'_3}{H_c-1}\vast).
\end{aligned}
\end{equation}
\normalsize

\subsection{Implementation of the ONC-Broadcast Scheme}\label{GS}
As clarified in the previous section, our proposed ONC-broadcast scheme can be implemented at the MBS  as follows:
\begin{enumerate}
\item \textbf{Build} the FCs' dual conflict graph as in Section \ref{HBSACG} and find the maximum independent set in this graph. The size of this maximum independent set is $N_{FC}$ vertices, which represent the requests to be served by the FCs.
\item \textbf{Remove} all the vertices corresponding to the $N_{FC}$  clients served by the FCs from the vertex set, and use the remaining vertices to build the MBS conflict graph. This graph will have $U-N_{FC}$ vertices.
\item \textbf{Determine} the minimum graph coloring of the MBS conflict graph. This represents the ONC solution at the MBS.
\item  \textbf{Compare} the ONC solution to the uncoded broadcasting solution $F-C$, and choose the one with the smaller size  as the MBS solution. This step is necessary to satisfy (\ref{Eq.201}) in the case when the remaining number of clients not served by the FCs $U-N_{FC}$ is greater than or equal to the remaining number of files $F-C$.
\end{enumerate}

Clearly, the aforementioned algorithm requires the solution of two NP-hard problems, namely the maximum independent set and minimum graph colouring problems. The former problem can be solved using well-known solvers, such as Bron-Kerbosch (B-K) algorithm \cite{BKMCFA}, for small networks settings. However, the complexity of B-K algorithm may still be prohibitive for real-time practical implementations in large network settings (i.e., with large number of FCs, clients and popular files). We thus propose the use of a simple greedy vertex search (GVS) algorithm in such practical settings. To solve using the GVS algorithm, we can assign to each vertex $v_{i,j,k}$ in the FCs' graph a weight $w_{i,j,k}$ defined as follows:

\small
\begin{equation}
w_{i,j,k} = (V -\delta_{i,j,k}) ~. \sum_{v_{i',j',k'}\in\mathcal{N}(v_{i,j,k})} (V-\delta_{i',j',k'})
\end{equation}
\normalsize

where $V$ is the total number of vertices in the graph, $\delta_{i,j,k}$ is the degree\footnote{The degree of a vertex in a graph is the number of vertices adjacent to this vertex.} of vertex $v_{i,j,k}$, and $\mathcal{N}(v_{i,j,k})$ is the set of its adjacent vertices. Thus, a vertex will have a high weight when it has a large number of non-adjacent vertices, which themselves have large numbers of non-adjacent vertices. We can then perform a maximum weight vertex search, by picking at each iteration the vertex with the maximum weight and then removing all its adjacent (i.e., conflicting) vertices from the graph before the next iteration. 

For the minimum graph coloring problem, we propose the use of greedy graph coloring (GGC), which is a very simple and  efficient graph coloring technique \cite[p.294]{random}. The greedy coloring algorithm is implemented as follows. The algorithm starts with the first vertex and adds it to the first color, then compares the second vertex to it. If the second vertex is not connected to the first vertex, it will be added to the first color too, otherwise, it will be given a new color, and so on until all vertices are colored. 

To summarize, the GVS algorithm is proposed to solve the maximum independent set problem at the FCs, and the GGC algorithm is suggested to find the minimum graph coloring at the MBS. We call this proposed heuristic the GVS-GGC algorithm. In the following parts, we comment on the performance and the complexity of this algorithm.

\subsection{Performance of the GVS-GGC Heuristic}
In this part, we discuss the performance of the proposed heuristic, in terms of the number of required MBS channels. We focus on the full FC coverage case to simplify the simulation in this section. Two bounds on the performance of the proposed heuristic are provided in the next theorem.

\begin{theorem}\label{Thm10}
Assuming full FC coverage, the performance of the GVS-GGC algorithm $N_{MBS}^{(\text{GVS-GGC})}$ satisfies

\small
\begin{equation}\label{Eq.400}
\begin{aligned}
&\min\left(\frac{\nu}{2\log_d (\nu)},F-C\right)\leq N_{MBS}^{(\text{GVS-GGC})} \leq \min\left(\left(1+\frac{5\log(\log (\nu))}{\log (\nu)}\right)\frac{\nu}{\log_d (\nu)},F-C\right),
\end{aligned}
\end{equation}
\normalsize

where $d=\frac{1}{1-\pi}$, $\nu$  and $\pi$ are the number of vertices and the conflict probability in the MBS graph, respectively.
\end{theorem}

\begin{IEEEproof}
The first term in the $\min()$ function of the lower bound is found by substituting (\ref{Eq.3}) in (\ref{Eq.201})  and setting $o_2(1)$  to zero. The first term inside the $\min()$ function of the upper bound was derived in \cite{McDiarmid}.
\end{IEEEproof}

\begin{figure}[!t]
\centering
\includegraphics[scale=.5, trim={0 7cm 0 7cm},clip]{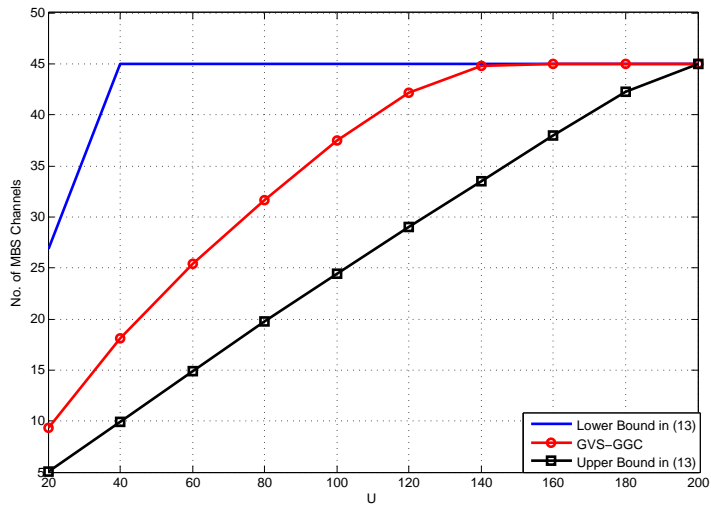}
\caption{The performance of the GVS-GGC heuristic compared to the upper and lower bounds in (\ref{Eq.400}). $F=50$, $C=5$, $\sigma_c=0.5$, $\sigma_u=0.2$.}\label{CB}
\end{figure}

Fig. \ref{CB} shows the performance of the GVS-GGC heuristic and the bounds in (\ref{Eq.500}). We observe that the three performance curves in Fig. \ref{CB} approach each other as they all converge to $F-C$ with the increase of the number of clients $U$.  Notice that the lower bound in (\ref{Eq.500}) is not necessarily achievable, since it represents the minimum chromatic number of the graph. Therefore, the performance of the GVS-GGC heuristic is expected to be closer to the actual best performance of the ONC-broadcast scheme, as will be shown in Section \ref{COGS}.

It is noteworthy to mention here that the bounds in (\ref{Eq.400}) can be generalized to the limited FC coverage case by averaging both sides of the inequality over the distribution of the number of vertices in the FC dual conflict graph $\tilde{\nu}$ the same way as in (\ref{Eq.40}).

\subsection{Worst-Case Time Complexity of the GVS-GGC Heuristic}\label{WCTCMWVSA}
To appreciate the simplicity of the proposed heuristic, we analyze its worst-case time complexity and compare it to that of the optimal ONC-broadcast scheme. First, we analyze the worst-case time complexity of the GVS algorithm. To start with the worst-case analysis, first we need to choose an operation that runs in a constant factor of time which is the most recurring in the algorithm. Here, we chose the operation of comparing the sum of two rows in the adjacency matrix, since the suggested algorithm is all about finding the row with maximum weight and then finding its neighbor with maximum weight and so on. A trivial worst-case is assumed, which is the case when the graph is fully connected. In this case, every two rows in the adjacency matrix would be compared to find the one with maximum weight. Therefore, starting with $V$ rows in the adjacency matrix of the dual conflict graph (the same as the number of vertices in the graph), finding the row with the maximum weight needs $V-1$ comparison operations. Next, the row with maximum weight is removed from the adjacency matrix and we are left with $V-1$, and the number of operations needed to find the maximum weight row is $V-2$ and so on. Thus, the total number of operations is equal to $\sum_{i=1}^{V-1} (V-i)=\frac{1}{2}(V^2+V)$. Hence, the worst-case time complexity of the GVS algorithm is $O(V^2)$. Using a similar approach, it can be easily shown that the worst-case time complexity of the GGC algorithm is $O(V^2)$, too. This indicates that the worst-case time complexity of GVS-GGC algorithm is $O(V^2)$.

We said previously in this section that the B-K algorithm can be used to find the maximum independent set in the FC graph. The worst-case time complexity of the B-K algorithm was shown to be $O(3^{V/3})$ \cite{Tomita200628}, which is exponential time complexity. The optimal graph coloring is applied to the MBS graph which has a considerably fewer number of vertices. Thus, the worst-case time complexity is determined by the B-K algorithm and is $O(3^{V/3})$. Hence, the proposed GVS-GGC algorithm achieves a significant reduction in the worst-case time complexity as compared to the ONC-broadcast scheme. 

\section{Simulation Results and Discussion}\label{SR}
In this section, simulation results are presented to evaluate the performance of the ONC-broadcast solution in terms of the average number of MBS channels and the offloading gain $OG$, defined as the percentage of saved MBS channels as a result of using FCs (i.e.,$ OG=\frac{\hat{N}_{MBS}-N_{MBS}}{\hat{N}_{MBS}}\times 100\%$, where $\hat{N}_{MBS}$ is the average required number of MBS channels assuming no FCs are deployed). The results are generated for a small network setting to compare the proposed  GVS-GGC heuristic with the optimal ONC-broadcast scheme, and for a large network setting to evaluate the performance of the GVS-GGC algorithm against the FC coverage radius for two different schemes assumed to be utilized at the FCs.

\subsection{Comparison with the Optimal ONC-Broadcast Scheme}\label{COGS}
The network example simulated in this section includes 2 FCs and a total of 10 files in the library. Every FC caches 7 files, and each client has 1 file. For the case of limited FC coverage, the coverage radius of every FC is 50m, and the coverage radius of the MBS is 60m. The optimal ONC-broadcast scheme is implemented using B-K algorithm at the FCs and optimal graph coloring at the MBS.

\begin{figure}[!t]
\centering
\includegraphics[scale=.5, trim={0 7cm 0 7cm},clip]{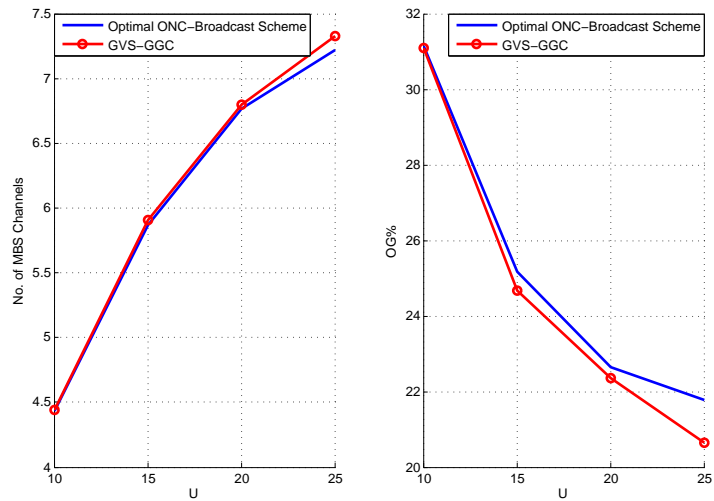}
\caption{The performance of the proposed GVS-GGC heursitic and the optimal ONC-broadcast scheme for the case of full FC coverage.}\label{AVPSFBKTH1}
\end{figure}

Fig. \ref{AVPSFBKTH1} plots the performance of both the GVS-GGC heuristic and the optimal ONC-broadcast scheme for the full FC coverage case. First, we observe that the performance achieved by the GVS-GGC is very close to that of the optimal ONC-broadcast scheme. We also note that, as a result of utilizing FCs, more than $20\%$ of the bandwidth required to serve all clients is offloaded from the MBS. 

The same observation can be made regarding the performance of the two schemes for the case of limited FC coverage. Furthermore, a degradation in the performance of both schemes can be observed for this case. This is expected since there is a set of clients not in the coverage set of each FC, and this limits the coding choices for each FC, and thus the offloading capabilities of the FCs. However, a significant offloading gain of at least $16\%$ can still be achieved.

\begin{figure}[!t]
\centering
\includegraphics[scale=.5, trim={0 7cm 0 7cm},clip]{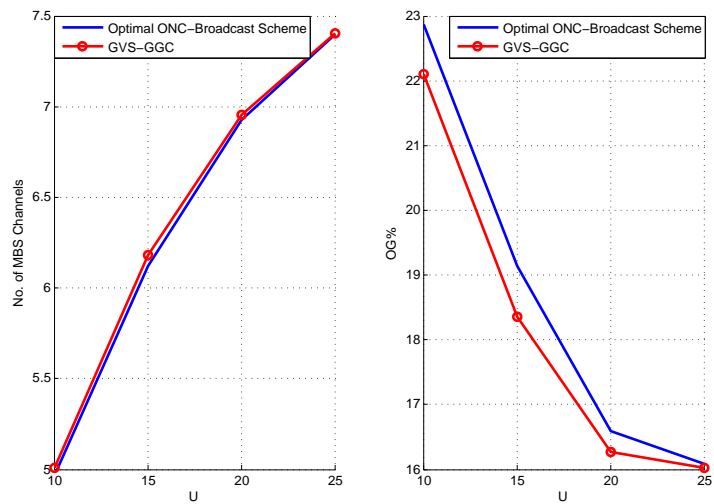}
\caption{The performance of the proposed heursitic and the optimal ONC-broadcast scheme for the case of limited FC coverage.}\label{AVPSFBKTH2}
\end{figure}

\subsection{Performance of GVS-GCC Versus the FC Coverage Radius}\label{CSGSLNS}
In this section, we conduct simulations to investigate the effect of the FC coverage radius on the performance of the GVS-GCC algorithm for a large network setting. We compare the results for the cases of using the dual conflict graph and the separate graph to solve at the FCs. The simulation comprises 32 FCs, 100 files. Each FC is assumed to cache 50 files, and each client has 10 files. The performance curves are plotted versus the coverage radius of each FC. The coverage radius of the MBS is set to 350m. 

\begin{figure}[!t]
\centering
\includegraphics[scale=.5, trim={0 7cm 0 7cm},clip]{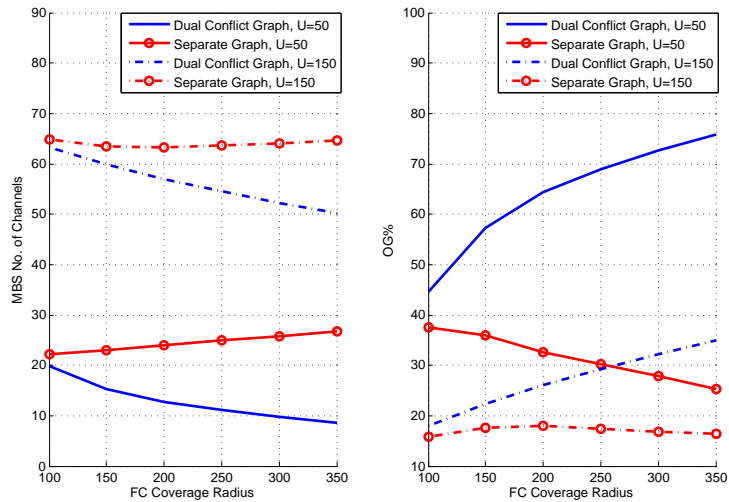}
\caption{The performance of the dual conflict and the separate graph schemes for the case of full FC coverage.}\label{LN1}
\end{figure}

Fig. \ref{LN1} depicts the designated comparison results for two values of the number of clients $U$. We observe that the performance of the dual conflict graph scheme is improved with the increase of the FC coverage radius, which can be justified by recalling that the dual conflict graph considers the transmission conflicts among the FCs. The number of transmission  conflicts between the vertices of the dual conflict graph increases with the number of clients in the intersection between the coverage areas of every two FCs, which increases with the increase of the coverage radius of each FC. On the other hand, the separate graph scheme shows either no improvement or worse performance with the increase of the FC coverage radius.  We can also observe that at least about  $45\%$ and $18\%$ of the bandwidth required to download all requested files is offloaded from the MBS when the dual conflict graph is used for $U=150$ and $U=50$, respectively.

\section{Conclusions}\label{CFW}
In this paper, the macrocell offloading problem in cellular networks is addressed. We first formulated the problem on a network coding graph and showed that it is an NP-hard problem. A ONC-broadcast scheme was then proposed to solve the problem by either maximizing the number of clients served by the FCs and then serving the remaining clients by the MBS or uncoded broadcasting at both the FCs and the MBS. The asymptotic optimality of the ONC-broadcast approach was shown, and a dual conflict graph and simple graph heuristics were developed to implement this ONC-broadcast approach. Moreover, the performance of the ONC-broadcast approach was analyzed in terms of the minimum achievable  number of orthogonal transmissions by the MBS for the case when systematic fixed placement is utilized to distribute the files among the FCs. Simulation results showed that the dual conflict graph scheme proposed for solving the content delivery problem at the FCs surpasses the intuitive separate graph scheme. Moreover, the  heuristic proposed to simplify the ONC-broadcast scheme was shown to achieve a close performance to that of the optimal greedy scheme. 

\section*{Appendix A\\ Proof of Theorem 3}\label{ApC}
Before we start with the proof, it is necessary to mention that the repetition index $R$ (defined as the number of copies of each file in the union of the caches of all FCs) is assumed to be integer in this section. To show that $N_{FC}^{(B)}$ is a lower bound on $N_{FC}$, we first assume a special case were $N_{FC}^{(B)}$ is maximum, namely, when all clients are in the coverage set of every FC. In this case, the number of clients served by the FCs in the broadcast case is equal to the average number of clients requesting a file multiplied by the number of FCs. Assuming that all files have the same probability to be chosen by a client, the number of clients requesting a file can modeled as a binomial random variable $\mathrm{Bin}(\tilde{U},\frac{1}{F})$, so the average number of clients requesting a file is equal to $\frac{U}{F}$. Thus, the average number of clients served by the FCs in the broadcast case is upper bounded by $\frac{CU}{F}$ (due to the FC full coverage assumption).

\begin{figure}[!t]
\centering
\includegraphics[scale=.5]{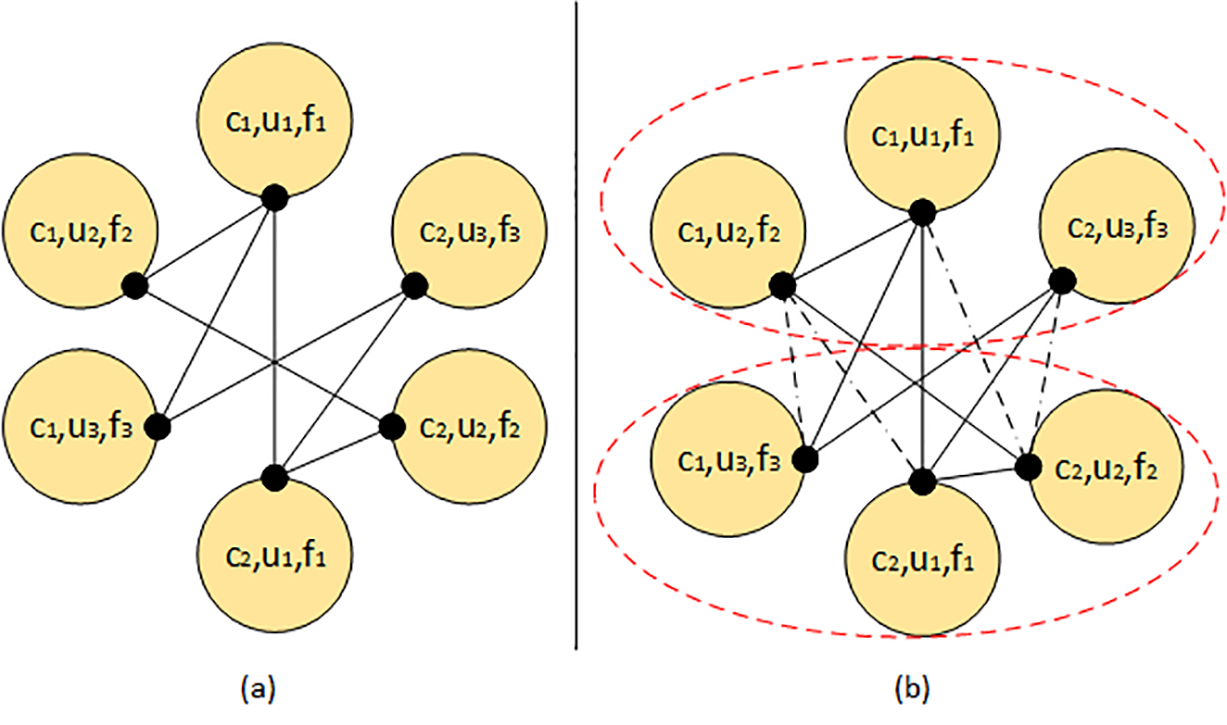}
\caption{An example showing (a) the original graph and (b) the new graph formed by connecting every vertex in the first subgraph to every vertex in the other subgraph. }\label{PRIL}
\end{figure}

To show that $\frac{CU}{F}\leq N_{FC}$, we form a new graph by decomposing the original FC graph into $R$ subgraphs (assuming $R$ is integer), and connecting each vertex in a subgraph to all vertices in all other subgraphs. The number of edges in the newly-formed FC graph is greater than that in the original FC graph since in the latter not every single vertex in a subgraph is connected to every vertex in all other subgraphs. To illustrate this fact, we take the following simple example comprising 3 files, 2 FCs and 3 clients. Each FC stores all packets, so $R=2$. The Has sets of clients $u_1$, $u_2$, and $u_3$ contain the files $f_3$, $f_3$ and $f_2$, respectively. The Wants sets of clients $u_1$, $u_2$, and $u_3$ contain the files $f_1$, $f_2$ and $f_3$, respectively. The original dual graph and the newly-formed graph are shown in Fig. \ref{PRIL}. The dash-dotted lines connecting the vertices in the new graph were added to make sure the average size of the solution of the new graph is $R N_{FC}^{(s)}$, where $N_{FC}^{(s)}$ is the average solution size of each subgraph. 

Apparently, the new graph has a greater number of edges than the original one, so the average size of the maximum independent set is smaller for the new graph (i.e., $R N_{FC}^{(s)}\leq N_{FC}$), and this is true for all possible ways of solving each subgraph. In particular, this is true if we solved each subgraph by broadcasting one uncoded file from each FC. Assuming that each FC has the same probability to induce a vertex in each subgraph which is equal to $\frac{1}{R}$, the average number of FCs in each subgraph is $\frac{C}{R}=\frac{F}{H_c}$. Since the average number of clients requesting a file is $\frac{U}{F}$, the average number of clients served in each subgraph in this case equals $\frac{F}{H_c}\times \frac{U}{F}=\frac{U}{F}$\footnote{We assume here that the number of clients requesting a file is independent from the number of FCs inducing a vertex in each subgraph, so the average of the multiplication of the two random variables is equal to the multiplication of the average of each one.}. Thus, 

\small
\begin{equation}\label{Eq.300}
R\frac{U}{H_c}\leq N_{FC}. 
\end{equation}
\normalsize

By substituting $R=\frac{H_c C}{F}$ in (\ref{Eq.300}), we get

\small
\begin{equation}\label{Eq.301}
R\frac{CU}{F}\leq N_{FC}. 
\end{equation}
\normalsize

To generalize, we argue that both $N_{FC}$ and $N_{FC}^{(B)}$ represent the size of a maximum independent set in general, which is a non-decreasing function on the number of vertices $\nu$, which is the same for the original and the new graphs as shown in the example in Fig. \ref{PRIL}. Thus, assuming that $F(\nu)=\frac{CU}{F}$ and $G(\nu)=N_{FC}$ (both denote the maximum assuming full FC coverage), we have just shown that $F(\nu)\leq G(\nu)$. Now, for the general case, the only difference in the original and the new graphs is the number of vertices which is reduced by a certain number (due to the removal of vertices induced by clients not included in the coverage set of an FC) which is the same for both graphs. Assuming that this number is $J$, we have in general $N_{FC}^{(B)}=F(\nu-J)$ and $N_{FC}=G(\nu-J)$, and hence $N_{FC}^{(B)}=F(\nu-J)\leq N_{FC}=G(\nu-J)$.

\section*{Appendix B\\ Proof of Lemma 1}\label{ApA}
For two vertices to be connected in the MBS conflict graph $\mathcal{G}^c$, two conditions are to be met
\small
\begin{enumerate}\label{cond_3}
\item $\mathrm{C1}$: $f_{k_1}\neq f_{k_2}$.
\item  $\mathrm{C2}$: $f_{k_1} \notin \mathcal{H}_{c_{j_2}}$ OR  $f_{k_2} \notin \mathcal{H}_{c_{j_1}}$.
\end{enumerate}
\normalsize
The first condition implies that the two vertices are induced by two different files, and the second condition implies that at least one of the clients inducing the two vertices does not have the file requested by the other client in its Has set. Clearly these two conditions are independent, so the vertex connectivity probability can be expressed as

\small
\begin{equation}\label{Eq.11}
\pi=\mathrm{Pr}\{\mathrm{C1}|\nu\}\mathrm{Pr}\{\mathrm{C2}|\nu\}=(1-\mathrm{Pr}\{\mathrm{\overline{C1}}|\nu\})(1-\mathrm{Pr}\{\mathrm{\overline{C2}}|\nu\}).
\end{equation}
\normalsize

$\mathrm{\overline{C1}}$ is the event that the two vertices of interest are induced by the same file. To find $\mathrm{Pr}\{\mathrm{\overline{C1}}\}$, we need to find the distribution of $\mathrm{\mathbf{z}}=[Z_1, ..., Z_F]$, where $Z_j$ is the number of clients requesting the $j^{th}$ file. Since the probability that a client would request a file is $\mu=\frac{1}{F}$, and a file in the MBS graph can be requested by any of the $\tilde{U}$ clients, $Z_j$ is the sum of $\tilde{U}$ Bernoulli trials, and thus can be modelled as a Binomial random variable $\mathrm{Bin}(\tilde{U},\frac{1}{F})$. Since the total number of vertices in  $\mathcal{\tilde{G}}$ is equal to the sum of the number of vertices induced by all files (i.e., $\nu=\sum_{v=1}^F Z_v$), we can say without loss of generality that $Z_F=\nu-\sum_{v=1}^{F-1} Z_v$. Therefore, given $\nu$, $\mathrm{\mathbf{z}}$ can be modelled as a multivariate hypergeometric distributed random vector \cite{Sorour}

\small
\begin{equation}\label{Eq.9}
\begin{aligned}
\mathrm{Pr}\{\mathrm{\mathbf{z}}= \mathrm{\mathbf{\hat{z}}}|\nu\}=\frac{\prod_{u=1}^{F-1}\dbinom{\tilde{U}}{z_u} \dbinom{\tilde{U}}{\nu-\sum_{v=1}^{F-1}z_v}}{\dbinom{\tilde{U}F}{\nu}},
\end{aligned}
\end{equation}
\normalsize
where $z_u\in\{0,1, \dots,\tilde{U}\}\; \forall u\in\{1, \dots, F-1\}$, and $\dbinom{.}{.}$ is the binomial coefficient.

Now, Since the vertices' identities are ignored, $\mathrm{Pr}\{\mathrm{\overline{C1}}\}$ can be written as

\small
\begin{equation}\label{Eq.12}
\mathrm{Pr}\{\mathrm{\overline{C1}}|\tilde{U},\mathrm{\mathbf{z}}= \mathrm{\mathbf{\hat{z}}}\}=\sum_{m=1}^{F}\frac{z_m(z_m-1)}{\nu(\nu-1)}.
\end{equation}
\normalsize

By deconditioning (\ref{Eq.12}) over $\mathrm{\mathbf{z}}$, $\mathrm{Pr}\{\mathrm{\overline{C1}}|\nu\}$ can be found as follows

\small
\begin{equation}\label{Eq.13}
\begin{split}
\mathrm{Pr}\{\mathrm{\overline{C1}}|\nu\}&=\mathbb{E}_{\mathrm{\mathbf{z}}|\nu}\left(\sum_{m=1}^{F}\frac{z_m(z_m-1)}{\nu(\nu-1)}\right) \\
&=\sum_{m=1}^{F}\mathbb{E}_{\mathrm{\mathbf{z}}|\nu}\left(\frac{z_m(z_m-1)}{\nu(\nu-1)}\right) \\
& =\sum_{m=1}^{F}\sum_{\substack{z_u=0\\ \forall u}}^{\tilde{U}}\frac{z_m(z_m-1)}{\nu(\nu-1)}\\
&\times \frac{\prod_{u=1}^{F-1}\dbinom{\tilde{U}}{z_u} \dbinom{\tilde{U}}{\nu-\sum_{v=1}^{F-1}z_v}}{\dbinom{\tilde{U}F}{\nu}}\\
&=\sum_{m=1}^{F}\frac{\tilde{U}(\tilde{U}-1)}{\tilde{U}F(\tilde{U}F-1)}\times \\
&\sum_{\substack{z_u=0\\ \forall u\neq m}}^{\tilde{U}}\sum_{z_m=2}^{\tilde{U}-2} \frac{\prod_{u=1}^{F-1}\dbinom{\tilde{U}}{z_u}\dbinom{\tilde{U}-2}{z_u-2} \dbinom{\tilde{U}}{\nu-\sum_{v=1}^{F-1}z_v}}{\dbinom{\tilde{U}F-2}{\nu-2}} \\
& =\sum_{m=1}^{F}\frac{(\tilde{U}-1)}{F(\tilde{U}F-1)}=\frac{(\tilde{U}-1)}{(\tilde{U}F-1)}.
\end{split}
\end{equation}
\normalsize

$\mathrm{\overline{C2}}$ requires both clients inducing the vertices to have the files requested by each other. The probability that a client has a file is $\sigma_u$. Therefore, $\mathrm{Pr}\{\mathrm{\overline{C2}}|\nu\}$ can be expressed as

\small
\begin{equation}\label{Eq.14}
\mathrm{Pr}\{\mathrm{\overline{C2}}|\nu\}=\sigma_u^2.
\end{equation}
\normalsize

The theorem follows from substituting (\ref{Eq.13}) and (\ref{Eq.14}) in (\ref{Eq.11}) and noticing that $\nu=\tilde{U}$ for the case of one request per client.


\section*{Appendix C\\ Proof of Theorem 2}\label{ApB}
The chromatic number of the MBS graph can be found using (\ref{Eq.3}) and (\ref{Eq.10}), and can be written as

\small
\begin{equation}\label{Eq.35}
\begin{aligned}
& \chi(U-N_{FC})= \left(\frac{1}{2}+o_2(1)\right)\times&\\
& \log\left(\frac{(U-N_{FC})F-1}{((U-N_{FC})F-1)-((1-\sigma_c^2)(U-N_{FC})(F-1))}\right)&\\
& \times \frac{U-N_{FC}}{\log(U-N_{FC})},
\end{aligned}
\end{equation}
\normalsize

where $N_{FC}$ is the average number of clients served by the FCs.

To find $N_{FC}$, we utilize a random model for the FCs' dual graph. Let $\mathcal{\tilde{G}}_{\tilde{\nu},\tilde{\pi}}$ be the FCs' random dual graph. First, the probability that two vertices are connected in $\mathcal{\tilde{G}}$ is found by defining the following events:\\
E1: The two vertices do not represent the request of the same file and at least one of the two clients inducing the vertices requests a file that is not in the has set of the other (coding conflict).\\
E2: The same client is served by two different FCs (service conflict).\\

The two aforementioned events are apparently mutually exclusive. Therefore, since two vertices are connected in $\mathcal{\tilde{G}}$ if either E1 or E2 is satisfied, the probability that two vertices are connected in $\mathcal{\tilde{G}}$ can be expressed as

\small
\begin{equation}\label{Eq.31}
\begin{aligned}
\tilde{\pi}=\mathrm{Pr}\{\mathrm{E1}|\tilde{\nu}\}+\mathrm{Pr}\{\mathrm{E2}|\tilde{\nu}\}.
\end{aligned}
\end{equation}
\normalsize

Since each client requests only one file, $\mathrm{E1}$ implies that the two vertices are induced by the same server. Thus, the probability that $\mathrm{E1}$ will occur is written as

\small
\begin{equation}\label{Eq.100}
\begin{aligned}
\mathrm{Pr}\{\mathrm{E1}|\tilde{\nu}\}&=\mathrm{Pr}\{\mathrm{1S}\cap \mathrm{C1}\cap \mathrm{C2}|\tilde{\nu}\}&\\
&=\mathrm{Pr}\{\mathrm{1S}|\tilde{\nu}\}\mathrm{Pr}\{\mathrm{C1}|\mathrm{1S},\tilde{\nu}\}\mathrm{Pr}\{\mathrm{C2}|\mathrm{1S},\mathrm{C1},\tilde{\nu}\},
\end{aligned}
\end{equation}
\normalsize

where $\mathrm{1S}$ is the event that the two vertices are induced by the same FC, $\mathrm{C1}$ is the event that the first coding conflict condition is satisfied (i.e., assuming that $k$ and $j$ are the files inducing the two vertices then $\mathrm{C1}=\{k\neq j\}$), and $\mathrm{C2}$ is the event that the second coding conflict condition is satisfied (i.e., assuming that $\mathcal{H}_{u_1}$ and $\mathcal{H}_{u_2}$ are the Has sets of the clients requesting $k$ and $j$, respectively, then $\mathrm{C2}=\{k\not\in \mathcal{H}_{u_2}\; \mathrm{OR} \; j\not\in \mathcal{H}_{u_1}\}$.

To find $\mathrm{Pr}\{\mathrm{1S}|\tilde{\nu}\}$, $X_i$ is defined as the number of vertices induced by FC $c_i$. Clearly, $X_i$ is the sum of $Y_i$ independent Bernoulli trials ($Y_i$ is the number of clients in the coverage set of FC $c_i$), and thus can be modelled as a binomial random variable $\mathrm{Bin}(Y_i,1/B)$, and this applies for all $i=1,\dots, C$. Since the total number of vertices in  $\mathcal{\tilde{G}}$ is equal to the sum of the number of vertices induced by all FCs (i.e., $\tilde{\nu}=\sum_{l=1}^C X_l$), we can say without loss of generality that $X_C=\tilde{\nu}-\sum_{l=1}^{C-1} X_l$. Given this fact and the number of clients in the coverage set of each FC, we can find the distribution of $\mathbf{x}=[X_1,\dots, X_C]$ as follows

\footnotesize
\begin{equation}\label{Eq.101}
\begin{aligned}
& \mathrm{Pr}\{\mathbf{x}=\mathbf{x}'|\sum_{l=1}^C X_l=\tilde{\nu}, \mathbf{y}=\mathbf{y}'\}=&\\
&\frac{\mathrm{Pr}\{\mathbf{x}=\mathbf{x}',\sum_{l=1}^C X_l=\tilde{\nu} | \mathbf{y}=\mathbf{y}'\}}{\mathrm{Pr}\{\sum_{l=1}^C X_l=\tilde{\nu} | \mathbf{y}=\mathbf{y}'\}}=&\\
&  \frac{\mathrm{Pr}\{X_1=x_1,\dots, X_{C-1}=x_{C-1}, X_C=\tilde{\nu}-\sum_{l=1}^{C-1} x_l|\mathbf{y}=\mathbf{y}'\}}{\mathrm{Pr}\{\sum_{l=1}^C X_l=\tilde{\nu} | \mathbf{y}=\mathbf{y}'\}}&\\
& = \prod_{v=1}^{C-1}\binom{y_v}{x_v} (1/B)^{x_v} (1-(1/B))^{y_v-x_v}\binom{y_C}{\tilde{\nu}-\sum_{l=1}^{C-1} x_l}&\\
&\times (1/B)^{\tilde{\nu}-\sum_{l=1}^{C-1} x_l}(1-(1/B))^{y_C-(\tilde{\nu}-\sum_{l=1}^{C-1} x_l)}&\\
& \times \left(\binom{\sum_{p=1}^C y_p}{\tilde{\nu}} (1/B)^{\tilde{\nu}} (1-(1/B))^{\sum_{p=1}^C y_p-\tilde{\nu}}\right)^{-1}&\\
&=\frac{\prod_{v=1}^{C-1} \binom{y_v}{x_v} \binom{y_C}{\tilde{\nu}-\sum_{l=1}^{C-1} x_l}}{\binom{\sum_{p=1}^C y_p}{\tilde{\nu}}}.
\end{aligned}
\end{equation}
\normalsize

Now, the probability that two vertices are induced by the same FC is written as

\small
\begin{equation}\label{Eq.102}
\begin{aligned}
\mathrm{Pr}\{\mathrm{1S}|\tilde{\nu},\mathbf{y}=\mathbf{y}',\mathbf{x}=\mathbf{x}'\}=\sum_{i=1}^C \frac{x_i(x_i-1)}{\tilde{\nu}(\tilde{\nu}-1)},
\end{aligned}
\end{equation}
\normalsize

where $x_C=\tilde{\nu}-\sum_{l=1}^{C-1} x_l$. Therefore, $\mathrm{Pr}\{\mathrm{1S}|\tilde{\nu},\mathbf{y}=\mathbf{y}'\}$ can be found as follows

\small
\begin{equation}\label{Eq.103}
\begin{aligned}
& \mathrm{Pr}\{\mathrm{1S}|\tilde{\nu},\mathbf{y}=\mathbf{y}'\}=\mathbb{E}_{\mathbf{x}|\tilde{\nu},\mathbf{y}}\left(\sum_{i=1}^C \frac{x_i(x_i-1)}{\tilde{\nu}(\tilde{\nu}-1)}\right)&\\
&=\sum_{i=1}^C \mathbb{E}_{\mathbf{x}|\tilde{\nu},\mathbf{y}}\left(\frac{x_i(x_i-1)}{\tilde{\nu}(\tilde{\nu}-1)}\right)&\\
&=\sum_{i=1}^C \sum_{\overset{x_u=0}{\forall u}}^{y_u} \frac{x_i(x_i-1)}{\tilde{\nu}(\tilde{\nu}-1)}\frac{\prod_{u=1}^{C-1} \binom{y_u}{x_u} \binom{y_C}{\tilde{\nu}-\sum_{l=1}^{C-1} x_l}}{\binom{\sum_{p=1}^C y_p}{\tilde{\nu}}}\\
&=\sum_{i=1}^C \frac{y_i(y_i-1)}{\sum_{p=1}^C y_p(\sum_{p=1}^C y_p-1)}\times &\\
&\sum_{\overset{x_u=0}{\forall u\neq i}}^{y_u} \sum_{x_i-2=0}^{y_i-2} \frac{\prod_{\overset{x_u=0}{u\neq i}}^{C-1} \binom{y_u}{x_u}\binom{y_i-2}{x_i-2} \binom{y_C}{\tilde{\nu}-\sum_{l=1}^{C-1} x_l}}{\binom{\left(\sum_{p=1}^C y_p\right)-2}{\tilde{\nu} -2}}&\\
&=\sum_{i=1}^C \frac{y_i(y_i-1)}{\sum_{p=1}^C y_p(\sum_{p=1}^C y_p-1)},
\end{aligned}
\end{equation}
\normalsize

where the last line results from the fact that the preceding one is a summation over all the sample space of a multivariate hypergeometric probability mass function (pmf).

Given that the two vertices are induced by FC $c_i$, the next step is to find $\mathrm{Pr}\{\mathrm{C1}|\mathrm{1S}, X_i=x_i, Y_i=y_i,\tilde{\nu}\}$. The number of vertices induced by file $j$, $Z_j$, can be modelled as a binomial random variable $\mathrm{Bin}(Y_i,\mu)$. Following the same steps as in Appendix A, the probability $\mathrm{Pr}\{\mathrm{C1}|1S, X_i=x_i, Y_i=y_i,\tilde{\nu}\}$ (where $X_i$ is the number of vertices induced by FC $c_i$) can be found, and is written as

\small
\begin{equation}\label{Eq.106}
\begin{aligned}
\mathrm{Pr}\{\mathrm{C1}|1S, X_i=x_i, Y_i=y_i,\tilde{\nu}\}=\frac{y_i (H_c-1)}{y_i H_c-1}.
\end{aligned}
\end{equation}
\normalsize

It can be noticed here that, since (\ref{Eq.106}) is not a function of $x_i$, deconditioning over the pmf of $x_i$ would result in the same expression.

\begin{figure}[!t]
\centering
\includegraphics[scale=.55]{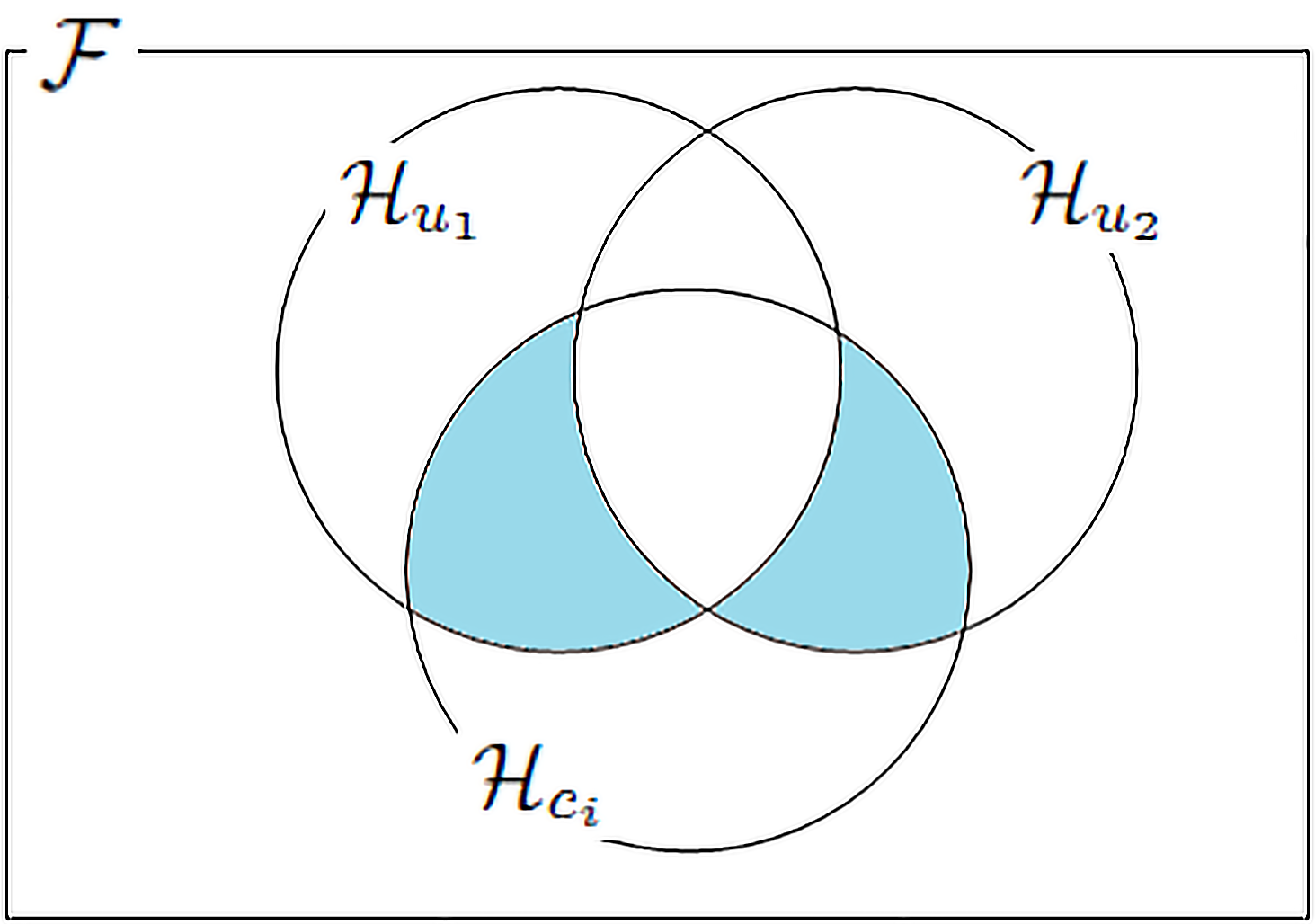}
\caption{An illustrative Venn diagram for the event $l\in \{\mathcal{H}_{u_1}-\mathcal{H}_{u_2}\}\cap \mathcal{H}_{c_i}\; \mathrm{AND}\; k\in \{\mathcal{H}_{u_2}-\mathcal{H}_{u_1}\}\cap \mathcal{H}_{c_i}$. }\label{IN1}
\end{figure}

Given the fact that the two vertices are induced by the same FC $c_i$ and two different files (file $k$ requested by client $u_1$ and file $l$ requested by client $u_2$), and knowing that file $k\not\in \mathcal{H}_{u_1}$ and file $l\not\in \mathcal{H}_{u_2}$, the event $\mathrm{\overline{C2}}$ is defined as $l\in \{\mathcal{H}_{u_1}-\mathcal{H}_{u_2}\}\cap \mathcal{H}_{c_i}\; \mathrm{AND}\; k\in \{\mathcal{H}_{u_2}-\mathcal{H}_{u_1}\}\cap \mathcal{H}_{c_i}$, and is illustrated in Fig. \ref{IN1} by the two shaded areas.

To find $\mathrm{Pr}\{\mathrm{\overline{C2}}| 1S,C1, Y_i=y_i, \tilde{\nu} \}$, we start at one of the shaded areas in Fig. \ref{IN1} and find the distribution of the number of files in it. Given this number, the distribution of the number of files in the second shaded area is found. Let us start with the shaded area in $\mathcal{H}_{u_1}$. The number of files in $\mathcal{H}_{u_1} \cap \mathcal{H}_{u_2}$, $e_1$, can be modelled using hypergeometric distribution as 

\small
\begin{equation}\label{Eq.108}
\begin{aligned}
\mathrm{Pr}\{e_1=e'_1 \}=\frac{\binom{H_u}{e'_1} \binom{F-H_u}{H_u-e'_1}}{\binom{F}{H_u}}.
\end{aligned}
\end{equation}
\normalsize

The first shaded area is defined as the set $\{\mathcal{H}_{u_1}-\mathcal{H}_{u_2}\}\cap \mathcal{H}_{c_i}$, and the number of files in this set $e_2$ can be modelled using hypergeomeric distribution given $e_1$ as 

\small
\begin{equation}\label{Eq.109}
\begin{aligned}
\mathrm{Pr}\{e_2=e'_2|e_1=e'_1 \}=\frac{\binom{H_c}{e'_2} \binom{F-H_c}{H_u-e'_1-e'_2}}{\binom{F}{H_u-e'_1}}.
\end{aligned}
\end{equation}
\normalsize

Thus now we know the distribution of the number of files in the first shaded area. Since it is given that $k$ would never be in the first shaded area, the second shaded area is defined as the set $\{\mathcal{H}_{u_2}-\mathcal{H}_{u_1}\}\cap \{\mathcal{H}_{c_i}-  \{\{\mathcal{H}_{u_1}-\mathcal{H}_{u_2}\}\cap \mathcal{H}_{c_i}  \}\}$, and given $e_1$ and $e_2$, the number of files in the second shaded area can be modelled using hypergeometric distribution as 

\small
\begin{equation}\label{Eq.110}
\begin{aligned}
\mathrm{Pr}\{e_3=e'_3|e_2=e'_2,e_1=e'_1 \}=\frac{\binom{H_c-e'_2}{e'_3} \binom{F-(H_c-e'_2)}{H_u-e'_1-e'_3}}{\binom{F}{H_u-e'_1}}.
\end{aligned}
\end{equation}
\normalsize

Hence, $\mathrm{Pr}\{\mathrm{\overline{C2}}|C1,1S,Y_i=y_i,\tilde{\tilde{\nu}}\}$ is determined as 

\small
\begin{equation}\label{Eq.111}
\begin{aligned}
& \mathrm{Pr}\{\mathrm{\overline{C2}}|C1,1S,Y_i=y_i,\tilde{\nu}\}=&\\
&\sum_{e'_1}\sum_{e'_2}\sum_{e'_3}\mathrm{Pr}\{e_3=e'_3,e_2=e'_2,e_1=e'_1 \}\frac{e'_2}{H_c}\frac{e'_3}{H_c-1}=&\\
& \sum_{e'_1}\sum_{e'_2}\sum_{e'_3}\mathrm{Pr}\{e_1=e'_1 \}\mathrm{Pr}\{e_2=e'_2|e_1=e'_1 \} \times &\\
& \mathrm{Pr}\{e_3=e'_3|e_2=e'_2,e_1=e'_1 \}\frac{e'_2}{H_c}\frac{e'_3}{H_c-1}=&\\
& \sum_{e'_1=0}^{H_u}\frac{\binom{H_u}{e'_1} \binom{F-H_u}{H_u-e'_1}}{\binom{F}{H_u}}\times &\\
& \sum_{e'_2=\max(0,H_u-(F-H_c))}^{\min(H_u-e'_1,H_c)}\frac{\binom{H_c}{e'_2} \binom{F-H_c}{H_u-e'_1-e'_2}}{\binom{F}{H_u-e'_1}}\times &\\
& \sum_{e'_3=\max(0,H_u-(F-(H_c-e'_2)))}^{min(H_u-e'_1,H_c-e'_2)} \frac{\binom{H_c-e'_2}{e'_3} \binom{F-(H_c-e'_2)}{H_u-e'_1-e'_3}}{\binom{F}{H_u-e'_1}} \times&\\
& \frac{e'_2}{H_c}\frac{e'_3}{H_c-1}.
\end{aligned}
\end{equation}
\normalsize

It should be noted here that (\ref{Eq.111}) is not a function of $y_i$, so deconditioning over it would result in the same expression.

Next, we need to find the pmf of $\mathbf{y}=[Y_1,\dots, Y_C]$ and decondition over it to get $\mathrm{Pr}\{\mathrm{E1}|\tilde{\nu}\}$. Since $Y_i$ is the number of clients in the coverage set of FC $c_i$, clearly $Y_i,\; i=1,\dots, C$ is the sum of $U$ independent Bernoulli trials and thus can be modelled as a binomial random variable $\mathrm{Bin}(U,P_{\mathcal{C}_i})$, where $P_{\mathcal{C}_i}$ is the probability that a client is in the coverage set of FC $c_i$ which, assuming the clients are uniformly distributed and all the FCs has the same coverage area, is the same for all FCs and equal to the coverage area of an FC divided by the total macrocell area. Since the event of interest is whether two vertices are connected in $\mathcal{\tilde{G}}$ (i.e., at least two vertices exist in the graph), the sum of the number of clients in all FCs' coverage sets should be at least two ($\sum_{p=1}^C Y_p\geq 2 $). Thus, the conditional pmf of $\mathbf{y}=[Y_1,\dots, Y_C]$ can be found as 

\small
\begin{equation}\label{Eq.104}
\begin{aligned}
& \mathrm{Pr}\{\mathbf{y}=\mathbf{y}'|\sum_{q=1}^C Y_q\geq 2\}=&\\
&\frac{\mathrm{Pr}\{Y_1=y_1,\dots, Y_{C-1}=y_{C-1}, Y_C\geq 2-\sum_{q=1}^{C-1} y_q\}}{\mathrm{Pr}\{\sum_{q=1}^C Y_q\geq 2\}}=&\\
&  \prod_{v=1}^{C-1}\binom{U}{y_v} (P_{\mathcal{C}})^{y_v} (1-P_{\mathcal{C}})^{U-y_v}\times &\\
&\sum_{y_C=\max(0,2-\sum_{q=1}^{C-1}y_q)}^U \binom{U}{y_C} P_{\mathcal{C}}^{y_C} (1-P_{\mathcal{C}})^{U-y_C}\times &\\
& \left(1-(1-P_{\mathcal{C}})^{UC-1}(P_{\mathcal{C}}(UC-1)+1)\right)^{-1},
\end{aligned}
\end{equation}
\normalsize

and, by deconditioning over $\mathbf{y}$, $\mathrm{Pr}\{\mathrm{E1}|\tilde{\nu}\}$ can be written as

\small
\begin{equation}\label{Eq.105}
\begin{aligned}
& \mathrm{Pr}\{\mathrm{E1}|\tilde{\nu}\}=&\\
& \sum_{\overset{y_v=0}{\forall v\neq C}}^U \sum_{y_C=\max(0,2-\sum_{q=1}^{C-1}y_q)}^U \sum_{m=1}^C \mathrm{Pr}\{\mathrm{1S}|\tilde{\nu},\mathbf{y}=\mathbf{y}'\}\times &\\
& \mathrm{Pr}\{\mathrm{C1}|1S, Y_i=y_i,\tilde{\nu}\} (1-\mathrm{Pr}\{\mathrm{\overline{C2}}|C1,1S,Y_i=y_i,\tilde{\nu}\})&\\
& =\sum_{\overset{y_v=0}{\forall v\neq C}}^U \sum_{y_C=\max(0,2-\sum_{q=1}^{C-1}y_q)}^U \sum_{m=1}^C \frac{y_m^2(y_m-1)}{\sum_{p=1}^C y_p(\sum_{p=1}^C y_p-1)}\times &\\
& \frac{(H_c-1)\prod_{v=1}^{C}\binom{U}{y_v} (P_{\mathcal{C}})^{y_v} (1-P_{\mathcal{C}})^{U-y_v}}{(y_m H_c-1)\left(1-(1-P_{\mathcal{C}})^{UC-1}(P_{\mathcal{C}}(UC-1)+1)\right) }  \vast(1- &\\
& \sum_{e'_1=0}^{H_u}\frac{\binom{H_u}{e'_1} \binom{F-H_u}{H_u-e'_1}}{\binom{F}{H_u}}\sum_{e'_2=\max(0,H_u-(F-H_c))}^{\min(H_u-e'_1,H_c)}\frac{\binom{H_c}{e'_2} \binom{F-H_c}{H_u-e'_1-e'_2}}{\binom{F}{H_u-e'_1}}\times  &\\
& \sum_{e'_3=\max(0,H_u-(F-(H_c-e'_2)))}^{min(H_u-e'_1,H_c-e'_2)} \frac{\binom{H_c-e'_2}{e'_3} \binom{F-(H_c-e'_2)}{H_u-e'_1-e'_3}}{\binom{F}{H_u-e'_1}} \frac{e'_2}{H_c}\frac{e'_3}{H_c-1}\vast).
\end{aligned}
\end{equation}
\normalsize

To find $\mathrm{Pr}\{\mathrm{E2}|\tilde{\nu}\}$, we define $L_i$ as the number of vertices induced by client $u_i$. Assuming that the events $u_i\in \mathcal{U}(c_j)$ and $u_i\in \mathcal{U}(c_k)$ are independent for all clients and for any arbitrary two FCs \footnote{This assumption is valid when the FCs are randomly dispersed, which is usually the case in two-tier networks due to the demand-based deployment of small cells \cite{Heath}. However, for the case of deploying the FCs in predetermined locations, as in the model studied in \cite{Shanmugam}, this assumption becomes valid for large coverage radii of the FCs.}, $L_i$ can be modelled using a binomial distribution $\mathrm{Bin}(R,\sigma_c P_{\mathcal{C}})$. Thus, $\mathrm{Pr}\{\mathbf{l}=\mathbf{l'}|\sum_{v=1}^U L_v=\tilde{\nu}\}$, where $\mathbf{l}=[L_1,\dots, L_U]$, is written as

\small
\begin{equation}\label{Eq.120}
\begin{aligned}
\mathrm{Pr}\{\mathbf{l}=\mathbf{l'}|\sum_{v=1}^U L_v=\tilde{\nu}\}=\frac{\prod_{u=1}^{U-1}\dbinom{R}{l_u} \dbinom{R}{\tilde{\nu}-\sum_{v=1}^{U-1}l_v}}{\dbinom{R U}{\tilde{\nu}}},
\end{aligned}
\end{equation}
\normalsize

and $\mathrm{Pr}\{\mathrm{E2}|\tilde{\nu} \}$ is found as 

\small
\begin{equation}\label{Eq.121}
\begin{aligned}
& \mathrm{Pr}\{\mathrm{E2}|\tilde{\nu}\}=\mathbb{E}_{\mathbf{l}|\tilde{\nu}}\left(\sum_{m=1}^U \frac{l_m(l_m-1)}{\tilde{\nu}(\tilde{\nu}-1)}\right)&\\
&=\sum_{m=1}^U \mathbb{E}_{\mathbf{l}|\tilde{\nu}}\left(\frac{l_m(l_m-1)}{\tilde{\nu}(\tilde{\nu}-1)}\right)&\\
&=\sum_{m=1}^U \sum_{\overset{l_u=0}{\forall u}}^{R} \frac{l_m(l_m-1)}{\tilde{\nu}(\tilde{\nu}-1)}\frac{\prod_{u=1}^{U-1} \binom{R}{l_u} \binom{R}{\tilde{\nu}-\sum_{q=1}^{U-1} l_q}}{\binom{UR}{\tilde{\nu}}}&\\
&=\sum_{m=1}^U \frac{R(R-1)}{UR(UR-1)}\times &\\
&\sum_{\overset{l_u=0}{\forall u\neq m}}^{R} \sum_{l_m-2=0}^{R-2} \frac{\prod_{\overset{l_u=0}{u\neq m}}^{U-1} \binom{R}{l_u}\binom{R-2}{l_m-2} \binom{R}{\tilde{\nu}-\sum_{q=1}^{U-1} l_q}}{\binom{UR-2}{\tilde{\tilde{\nu}} -2}}&\\
&=\frac{R-1}{UR-1}.
\end{aligned}
\end{equation}
\normalsize

Given $\tilde{\nu}$, $N_{FC}$ can be determined by finding the size of the maximum independent set in $\mathcal{\tilde{G}}$, which is the same as the clique number of the graph. The clique number of $\mathcal{\tilde{G}}$ ($cl(\mathcal{\tilde{G}})$) can be approximated as \cite[pp. 283]{random}

\small
\begin{equation}\label{Eq.34}
\begin{aligned}
cl(\mathcal{\tilde{G}})&=2\log_{b}(\tilde{\nu})-2\log_b(\log_b(\tilde{\nu}))+2log_b(\mathrm{e}/2)&\\
&+1+o_1(1),
\end{aligned}
\end{equation}
\normalsize

where $b=\frac{1}{1-\tilde{\pi}}$, and $o_1(1)$ is a number between 0 and 1 that can be found empirically. By  simplifying (\ref{Eq.34}), $N_{FC}$ can be written as

\small
\begin{equation}\label{Eq.41}
\begin{aligned}
N_{FC|\tilde{\nu}}=2\log_{\frac{1}{1-\tilde{\pi}}}\left(\frac{e\tilde{\nu}}{2\log_{\frac{1}{1-\tilde{\pi}}}(\tilde{\nu})}\right)+1+o_1(1).
\end{aligned}
\end{equation}
\normalsize

Finally, the distribution of $\tilde{\nu}$ can be found by realizing that $\tilde{\nu}=\sum_{n=1}^U L_n$. So $\tilde{\nu}$ can be modelled using the distribution of the sum of independent binomial random variables, which is a binomial distribution. However, since the number of vertices should be at least 2, $\tilde{\nu}$ is modelled using a truncated binomial distribution, and thus $\mathrm{Pr}\{\tilde{\nu}=\tilde{\nu}'|\tilde{\nu}\geq 2\}$ can be written as

\small
\begin{equation}\label{Eq.122}
\begin{aligned}
& \mathrm{Pr}\{\tilde{\nu}=\tilde{\nu}'|\tilde{\nu}\geq 2\}=&\\
& \frac{\binom{RU}{\tilde{\nu}'}(\sigma_c P_{\mathcal{C}})^{\tilde{\nu}'}(1-\sigma_c P_{\mathcal{C}})^{RU-\tilde{\nu}'}}{1-(1-\sigma_c P_{\mathcal{C}})^{RU-1}(\sigma_c P_{\mathcal{C}}(RU-1)+1)}.
\end{aligned}
\end{equation}
\normalsize

The theorem follows from substituting (\ref{Eq.41}) in (\ref{Eq.35}) and deconditioning over $\tilde{\nu}$.

\end{document}